\documentclass[a4paper,fleqn,usenatbib]{mnras}

\usepackage[T1]{fontenc}
\usepackage[utf8]{inputenc}
\usepackage{ae,aecompl}

\usepackage{graphicx}	
\usepackage{amsmath}	
\usepackage{amssymb}	
\usepackage{float} 
\usepackage[normalem]{ulem}
\def\mearth{{\rm\,M_\oplus}}

\title[Breaking the resonant chains]{Breaking the Chains: Hot Super-Earth systems from migration and disruption of compact resonant chains}

\author[A. Izidoro et al.]{
Andre Izidoro,$^{1,2}$\thanks{E-mail:izidoro.costa@gmail.com} Masahiro Ogihara$^{3}$, 
Sean N. Raymond,$^{1}$ Alessandro Morbidelli$^{4}$,  \newauthor Arnaud Pierens,$^{1}$
Bertram Bitsch,$^{5}$ Christophe Cossou $^{6}$ and Franck Hersant$^{1}$ 
\\
$^{1}$ Laboratoire d'astrophysique de Bordeaux, Univ. Bordeaux, CNRS, B18N, allée Geoffroy Saint-Hilaire, 33615 Pessac, France \\
$^{2}$ UNESP, Univ. Estadual Paulista - Grupo de Din{\^a}mica Orbital \& Planetologia, Guaratinguet{\'a}, CEP 12516-410 S{\~a}o Paulo, Brazil \\
$^{3}$ Division of Theoretical Astronomy, National Astronomical Observatory of Japan, 2-21-1, Osawa, Mitaka, Tokyo 181-8588, Japan \\
$^{4}$ Laboratoire Lagrange, UMR7293, Université Côte d’Azur, CNRS, Observatoire de la Côte d’Azur,\\ Boulevard de l’Observatoire, 06304 Nice Cedex 4, France \\
$^{5}$  Lund Observatory, Department of Astronomy and Theoretical Physics, Lund University, Box 43, 22100 Lund, Sweden \\
$^{6}$ Institut d'Astrophysique Spatiale, France
}

\date{Accepted XXX. Received YYY; in original form ZZZ}

\pubyear{2016}

\begin{document}
\label{firstpage}
\pagerange{\pageref{firstpage}--\pageref{lastpage}}
\maketitle

\begin{abstract}

``Hot super-Earths'' (or ``Mini-Neptunes'') between 1 and 4 times Earth's size with period shorter than 100 days orbit 30-50\% of Sun-like type stars. Their orbital configuration -- measured as the period ratio distribution of adjacent planets in multi-planet systems -- is a strong constraint for formation models. Here we use N-body simulations with synthetic forces from an underlying evolving gaseous disk to model the formation and long-term dynamical evolution of super-Earth systems. While the gas disk is present, planetary embryos grow and migrate inward to form a resonant chain anchored at the inner edge of the disk. These resonant chains are far more compact than the observed super-Earth systems. Once the gas dissipates resonant chains may become dynamically unstable.  They undergo a phase of giant impacts that spreads the systems out.  Disk turbulence has no measurable effect on the outcome.  Our simulations match observations if a small fraction of resonant chains remain stable, while most super-Earths undergo a late dynamical instability. Our statistical analysis restricts the contribution of stable systems to less than $25\%$. Our results  also suggest that the large fraction of observed single planet systems does not necessarily imply any dichotomy in the architecture of planetary systems. Finally, we use the low abundance of resonances in Kepler data to argue that, in reality, the survival of resonant chains happens likely only in $\sim 5\%$ of the cases.  This leads to a mystery: in our simulations only 50-60\% of resonant chains became unstable whereas at least 75\% (and probably 90-95\%) must be unstable to match observations.


\end{abstract}




\section{Introduction}

Among the thousands of confirmed exoplanets, hot super-Earths or mini-Neptunes -- with radii between 1 and ${\rm 4R_{\oplus}}$ ({\rm 1 $< M_{\oplus}$< 20}), orbiting very close to their host stars -- form by far the largest population \citep{mayoretal11,boruckietal10,boruckietal11,lissaueretal11b,schneideretal11,howardetal12,fabryckyetal12,petiguraetal13,howardetal13,dongzhu13,batalhaetal13,howardetal13,fressinetal13,mullallyetal15}.  Statistical studies suggest that about 1 out of  3 solar type stars (FGK spectral types) host a super-Earth with orbital period shorter than 100 days \citep{mayoretal11,howardetal12,fressinetal13,petiguraetal13}. Yet, close-in super Earths are often found in compact multi-planet systems (e.g. \cite{lissaueretal11a}). Their eccentricities and mutual orbital inclinations  are estimated to statistically concentrate around low and moderate values ({\it e~}$\lesssim$~0.1-0.2;~{\it i~}$\lesssim$~5~deg; \cite{mayoretal11,lissaueretal11a,fangmargot12}). 

A fundamental open question in planet formation is: {\it Where and how did hot super-Earth systems form and dynamically evolve?}



Current models  on the formation of systems of close-in super-Earths can be divided into two main categories~\citep[for reviews see][]{raymondetal08b,raymondetal14,morbyraymond16}: (1) in-situ accretion \citep{raymondetal08b,hansenmurray12,hansenmurray13,chianglaughlin13,hansen14,ogiharaetal15a,ogiharaetal15b} or (2) assembly of planets at moderate or larger distances from the star followed by inward gas-driven migration \citep{terquempapaloizou07,idalin08,idalin10,mcneilnelson10,hellarynelson12,cossouetal14,colemannelson14,colemannelson16}. In-situ accretion models virtually comes in two flavors: (a) standard in-situ accretion models which invokes high mass disks from the beginning to allow multi-Earth masses planets to form in the inner regions \citep{hansenmurray12,hansenmurray13}; and (b) drift-then-assembly model which proposes that these planets form in the innermost regions of the disk in consequence of a local concentration of pebbles or small planetesimals drifting inward due to gas drag \citep{chatterjeetan14,boleyford13,boleyetal14,chatterjeetan15,huetal16}.

The standard in-situ accretion model suffers from too many fundamental issues to be plausible (see discussions in \cite{raymondcossou14}, \cite{schlichting14}, \cite{schlaufman14}, \cite{izidoroetal15a}, \cite{ogiharaetal15a}, and \cite{chatterjeetan15}). For example, it ignores the effects of planet-disk gravitational interaction. Planets forming in-situ grow extremely fast because of short dynamical timescales and the required abundant amount of mass in the inner regions of the disk \citep{hansenmurray12,hansenmurray13}. They tend to reach masses large enough \citep{hansenmurray13,hansen14} to sufficiently perturb the surrounding gas in timescales much shorter than the expected lifetime of protoplanetary disks around young stars \citep{ogiharaetal15a}. Planet-disk gravitational interaction  leads to angular momentum transfer between the planet and the disk (see recent review by \cite{baruteauetal14}) which typically causes orbital radial migration\citep{goldreichtremaine79,goldreichtremaine80,linpapaloizou79,linpaploizou86,ward86,artymowicz93,ward97a,tanakaetal02,sarigoldreich04}, as well as eccentricity and inclination damping of the planets' orbit \citep{papaloizoularwood00,goldreichsari03,tanakaward04}. Ignoring planet-disk gravitational interaction  (migration and orbital tidal damping) in models of in-situ formation is not self-consistent. Moreover, if planets eventually migrate during the gas disk phase they would not be forming truly in-situ \citep{ogiharaetal15a}. 

 The drift-then-assembly models \citep{chatterjeetan14,boleyford13,boleyetal14,chatterjeetan15,huetal16} are quite promising but what actually happens near the inner edge of the disk, which is dangerously close to the sublimation line of silicates is not clear \citep{morbidellietal16}. Instead, the formation of massive objects beyond the snowline within the lifetime of the disk seems to be generic from theoretical considerations \citep[e.g.][]{morbidellietal15b}. Migration seems to be a generic process as well. So, we focus in this paper on the scenario where planets are assembled at moderate or larger distances from the star and then moved close to the star by gas-driven migration.

Earth-mass planets typically are not able to open a gap in the gaseous disk (e.g. \cite{papaloizoulin84,cridaetal06}) and migrate in the type-I regime (e.g. \cite{ward97a,kleynelson12}). Sophisticated hydrodynamical simulations including thermodynamical  effects show that type-I migration is very sensitive to the disk properties. Planets may either migrate inward or outward depending on the combination of different torques from the disk \citep{paardekoopermellema06,paardekoopermellema08,baruteaumasset08,paardekooperpapaloizou08,kleyetal09}. Indeed, the Lindblad torque tend to push the planet inwards (e.g. \cite{ward86,ward97a}) but depending on the planet mass the gas-flowing in coorbital motion with the planet may exert a strong torque capable of stopping or even reversing the direction of type-I migration (\cite{kleycrida08,paardekooperetal10,paardekooperetal11}.  There are locations within the disk where the net torque is zero (Lyra et al 2010; Horn et al 2012; Cossou et al 2013; Pierens et al 2013; Bitsch et al 2013, 2014, 2015).  However, as the disk dissipates and cools,  thermodynamics and viscous effects becomes less important and planets are released to migrate inward \citep{lyraetal10,hornetal12,bitschetal14}. It is hard to imagine planets completely escaping inward type-I migration, although disk winds may be a possible explanation of a global  suppression of type-I migration in the inner parts of the disk \citep{ogiharaetal15a,suzukietal16}.

One of the main criticisms of the inward migration model for the origins of close-in super-Earths comes from the fact that many planet pairs are near but not exactly in first order mean motion resonances (\cite{lissaueretal11b,fabryckyetal14}). There is a prominent excess of planet pairs just outside first order mean motion resonances \citep{fabryckyetal14}. Planet migration models predicts that as the disk dissipates planets should migrate inward and pile up in long chains of mean motion resonances (\cite{terquempapaloizou07,raymondetal08b,mcneilnelson10,rein12,reinetal12,hornetal12,ogiharakobayashi13,haghighipour13,cossouetal14,raymondcossou14,ogiharaetal15a,liuetal15,liuetal16}), in stark contrast to the observations. There are three reasons not to reject the migration model. First, a fraction of planet pairs are indeed in first order mean motion resonance (\cite{lissaueretal11b,fabryckyetal14}). For example, the recently discovered Kepler 223 system presents a very peculiar orbital configuration. Planets in this system are locked in a chain of resonances which mostly likely could be explained by convergent migration during the gas disk phase \citep{millsetal16}. The TRAPPIST-1 system is another  example of planetary system with multiple planets in a resonant chain \citep{gillonetal17}. Second,  a number of mechanisms have been proposed for accounting for the excess of planet pairs outside mean motions resonances which could consistently operate with the inward-migration model. Among them are star-planet tidal dissipation \citep{papaloizouterquem10,papaloizou11,lithwickwu12,delisleetal12,batyginmorbidelli13b,delisleetal14,delislelaskar14},  planet scattering of leftover planetesimals \citep{chatterjeeford15},  turbulence in the gaseous disk \citep{rein12,reinetal12} (but see Section 4), interaction with wake excited by other planets \citep{baruteaupapaloizou13} and the effects of asymmetries in the structure of the protoplanetary disk \citep{batygin15}. Third, and most importantly, planets that form in resonance may not remain in resonance.  Rather, planets in resonance can become unstable when the gas disk dissipates \citep{terquempapaloizou07,ogiharaida09,cossouetal14}.  The systems that survive instabilities are not in resonance. To summarize, the lack of resonance between planet pairs should not be taken as evidence against inward-migration \citep{goldreichschlichting14}.  
 
In this paper we assume that hot super-Earth systems form by  inward gas-driven migration. Our numerical simulations model the dynamical evolution of Earth-mass planets in evolving protoplanetary disks \citep{willianscieza11}. Our nominal simulations include the effects of type-I migration and orbital eccentricity and inclination damping due to the gravitational interaction with the  gas. We have also tested in our model the effects of stochastic forcing from turbulent fluctuations in the disk of gas. After the gas disk's dissipation, our simulations were continued up to 100~Myr. The goal of this study is to help elucidating the following question: is the inward migration model for the origins of hot super-Earths systems consistent with observations? 

This paper is structured as follows. In Section 2 we describe our methods and disk model. In Section 3 we present the results of simulations of our fiducial model. In Section 4 we describe our turbulent model and the results of simulations including these effects.   In Section 5 we discuss about the role of dynamical instabilities post-gas disk dissipation and the final dynamical architecture of planetary systems produced in our simulations. In section 6 we compare the results of our fiducial and turbulent models with observations. In Section 7 we compare our results with other models in the literature. In Section 8 we discuss about our main results. Finally, in Section 9 we summarize our conclusions.


\section{Methods}
We use N-body numerical simulations to study the dynamical evolution of multiple Earth-mass planets in evolving protoplanetary disks. We  also follow the subsequent phase of dynamical evolution of formed planetary systems post gas-disk dissipation. During the gaseous phase, we mimic the effects of the disk of gas on the planets by applying artificial forces onto the planets (or protoplanetary embryos). These forces were calibrated from truly hydrodynamical simulations. In this section we describe our gas disk model, followed by the details of our prescription for type-I migration, eccentricity and orbital inclination damping, and finally we explain how we set the initial distribution of protoplanetary embryos in the system. We also performed simulations testing the effects of stochastic forcing from turbulent fluctuations in the disk. 

To perform our simulations we use an adapted version of Mercury~\citep{chambers99}. In all our simulations, collisions are considered perfect merging events that conserve linear momentum. During the gas disk phase, our simulations adapted the global timestep to reduce the integration time. Every 1000 timesteps, the timestep was re-evaluated and, if necessary, decreased to be at most 1/25th of the orbital period at the perihelion of the innermost planet. While this technique is not strictly symplectic, we saw no difference in outcome when using it (although it significantly sped up the simulations).

\subsection{Disk-Model}

The initial structure of a protoplanetary disk can be derived from the radial disk temperature, the gas surface density and viscosity profiles. To model the disk's structure and evolution we incorporated the 1D disk model fits derived by \cite{bitschetal15} into our numerical integrator. There are two major advantages in using this  approach rather than calculating the evolution of a viscous disk. The first one is that these fits have been calibrated from sophisticated 3D hydrodynamical numerical simulations including effects of viscous heating, stellar irradiation and radial diffusion. The second one is that this approach is computationally cheaper --and for the purposes of this work-- more versatile and robust than having to solve a 1D disk evolution model to account for the disk evolution (e.g. \cite{colemannelson14}).

From the standard parametrized accretion rate on the star, given by
\begin{equation}
\dot{M}_\text{gas} =  3\pi\alpha h^2 r^2 \Omega_k \Sigma_{gas},
\end{equation}
and the hydrostatic equilibrium equation 
\begin{equation}
T =  h^2 \frac{G M_{\odot}}{r}\frac{\mu}{\mathcal{R}}
\end{equation}
it is straightforward to determine the disk surface density $\Sigma_{gas}$ using the disk temperature profile given in \cite{bitschetal15}. In Eq. (1) $\alpha$ is the dimensionless ${\rm \alpha}$-viscosity \citep{shakurasunyaev73}, h is the disk aspect ratio, r is the heliocentric distance, and ${ \Omega_k = \sqrt{GM_{\odot}/r^3}}$ is the keplerian frequency. T is the disk temperature at the midplane, G is the gravitational constant, ${ M_{\odot}}$ is the stellar mass, and ${\mu}$ is the mean molecular weight. In all our simulations the central star is one Solar mass and ${\rm \mu=2.3gmol^{-1}}$. The disk age (or alternatively ${\rm \dot{M}_{gas}}$) is approximated by the follow equation from  \cite{hartmannetal98} and modified by \cite{bitschetal15}
\begin{equation}
{log{\left( \frac{\dot{M}_{gas}}{M_{\odot}/yr}\right) }~=~-8 - 1.4log\left(\frac{t_{disk} + 10^5~yr}{10^6~yr}\right)}.
\end{equation}

We do not recalculate the disk structure following the fits in \cite{bitschetal15} at every timestep of the numerical integrator. Instead, we solve the disk structure every 500 years. Since the disk structure changes on a longer timescale this approach does not affect the validity of our conclusions and allows us to save substantial computational time. 

Using Eq. 1-3 we determine the disk temperature using the temperature profiles fits from the Appendix of \cite{bitschetal15}  which are given for different disk metallicities and different regimes of accretion onto the star (or ages of the disk). Following \cite{bitschetal15} our disk opacity is the same used in \cite{belllin94}. In our fiducial simulations the disk metallicity is set 1\% and the disk ${\rm \alpha}$-viscosity is set $\alpha = 5.4 \times 10^{-3}$. In this work we do not explore the effects of these parameters. However, our disk presents all main characteristics expected for a proto-planetary disk, with an inner edge and temporary outward migration zones. None of the results that we will obtain will be dependent on specific characteristics of this disk (e.g. the specific location of an outward migration zone), so we expect that they are fairly robust. The issue of the disk's lifetime will be discussed in Section 5.

The gas disk lifetime in our simulations is set to 5.1 Myr. As discussed in \citep{bitschetal15}, after the gas accretion rate onto the star drops  below $10^{-9}$ the gas density becomes so low that the disk can be evaporated in very short timescales. Thus, as stressed in  \cite{bitschetal15}  these fits should not be used to track the disk evolution beyond disk ages corresponding to ${\rm \dot{M}_{gas} = 10^{-9}~M_{\odot}/yr}$. In our simulations, we allow the disk evolve from $t_{disk} = 0$ to 5 Myr (${\rm \dot{M}_{gas} = 10^{-9}}~M_{\odot}/yr$ ) then we freeze the disk structure at 5 Myr and we exponentially decrease the surface density using an e-folding timescale of 10 Kyr. After 100 Kyr (at  $t_{disk}=5.1$ Myr) the disk is assumed to instantaneously dissipate. This allows a smooth transition from the gas-disk to the gas-free phase. 

 As the disk gets older and thinner low mass planets may eventually be able to open a gap in the disk \citep[e.g.][]{cridaetal06}. In our simulations, at the very late stages of the disk the disk's aspect ratio is about $\sim$0.03 near 0.1 AU. For a 10 Earth mass planet in the very inner regions of the disk the gap should not be fully open yet. Thus, for simplicity we have neglected effects of type-II migration in our simulations.

 Magnetohydrodynamic simulations of disk-star interaction suggest that a young star with sufficiently strong dipole magnetic field is surrounded by a low-density gas cavity \citep{romanovaetal03,flocketal17,bouvieretal07}. In this context, the inner edge of the  circumstellar disk typically corresponds to the approximate location where the angular velocity of the star equals to the keplerian orbital velocity, and migration should not continue within the cavity except in unusual circumstances \citep{romanovaetal06}. As the stellar spin rate evolves, the location of the inner edge would evolve as well. The observed stellar spin rate is between 1-10 days \citep[e.g.][]{bouvieretal13}, which suggest that the corotation radius is 0.01-0.2AU. The orbital period distribution of the innermost Kepler planets is consistent with this. We have included this characteristic of disks in our simulations considering that fixing the inner edge of the disk  at 0.1~AU is reasonable. In all our simulations the disk extends from 0.1 to 100 AU. At the inner edge of the disk, the surface density is artificially changed to create a planet trap at about $\sim$0.1 AU. This is done by multiplying the surface density by the following rescaling factor:

\begin{equation}
\Re = \tanh\left( \frac{r-0.1}{0.005}\right).
\end{equation}

\subsection{Disk-planet interaction: Type-I migration}

Based on the underlying disk profile we calculate type-I migration, eccentricity and orbital eccentricity damping. Our simulations start with planets that migrate in the type-I regime. The negative of the surface density profile and temperature gradients are given by

\begin{equation}
{ x = - \frac{\partial ln ~\Sigma_{gas}}{\partial ln~r},~~~\beta = - \frac{\partial ln ~T}{\partial ln~r} }.
\end{equation}

Following \citep{paardekooperetal10,paardekooperetal11} and assuming a gravitational smoothing length for the planet potential of ${\rm b=0.4h}$, the total torque from the gas experienced by a type-I migrating planet can be expressed by

\begin{equation}
{\Gamma_\text{tot} = \Gamma_\text{L}\Delta_\text{L} + \Gamma_\text{C}\Delta_\text{C}},
\end{equation}
where ${\rm \Gamma_L}$ is the Lindblad torque and ${\rm \Gamma_C}$ is the corotation torque from the gravitational interaction of the planet with the gas flowing around its orbit. The total torque that a planet feels also depends on its orbital eccentricity and inclination \citep{bitschkley10,bitschkley11,cossouetal13,pierensetal13,fendykenelson14}. To account for this we calculate ${\rm \Gamma_\text{tot}}$ with two rescaling functions to reduce the Lindblad and corotation torques accordingly to the planet's eccentricity and orbital inclination \citep{cresswellnelson08,colemannelson14}. The reduction of the Lindblad torque can be expressed as


\begin{multline}
 \Delta_\text{L}  = \left[   P_\text{e} + \frac{P_\text{e}}{|P_\text{e}|} \times \left\lbrace 0.07 \left( \frac{i}{h}\right)  + 0.085\left( \frac{i}{h}\right)^4 \right. \right. \\  \left. \left. - 0.08\left(  \frac{e}{h} \right) \left( \frac{i}{h} \right)^2 \right \rbrace \right] ^{-1}  ,
\end{multline}
where
\begin{equation}
{P_\text{e} = \frac{1+\left( \frac{e}{2.25h}\right)^{1.2} +\left( \frac{e}{2.84h}\right)^6}{1-\left( \frac{e}{2.02h}\right)^4}}.
\end{equation}

The reduction factor of the coorbital torque ${\rm \Delta_{C}}$ is simply given by

\begin{equation}
{ \Delta_\text{C}=\exp\left(\frac{e}{e_\text{f}} \right)\left\lbrace 1-\tanh\left(\frac{i}{h} \right)\right\rbrace   },
\end{equation}
where \textit{e} and \textit{i} are the planet orbital eccentricity and inclination, respectively. ${e_\text{f}}$ is  defined in \cite{fendykenelson14} as
\begin{equation}
{e_\text{f} = 0.5h + 0.01}.
\end{equation}

Under the effects of thermal and viscous diffusion the co-orbital torque  is written as:
\begin{align}
\Gamma_{\text{C}} ~=~\Gamma_{\text{c,hs,baro}} F(p_{\rm{\nu}}) G(p_{\rm\nu}) +  (1 - K(p_{\rm\nu}))\Gamma_{\text{c,lin,baro}} ~~+  \nonumber\\   \Gamma_{\text{c,hs,ent}}F(p_{{\rm \nu}})F(p_{\rm \chi})\sqrt{G(p_{\rm \nu})G(p_{\rm \chi})} \nonumber  +  \\ \sqrt{(1 - K(p_{\rm \nu}))(1 - K(p_{\rm \chi})}\Gamma_{\text{c,lin,ent}}~.
\end{align}


The formulae for ${\Gamma_\text{L}}$, ${ \Gamma_{\text{c,hs,baro}}}$, ${ \Gamma_{\text{c,lin,baro}}}$, ${\Gamma_{\text{c,hs,ent}}}$, and ${\Gamma_{\text{c,lin,ent}}}$ are
\begin{equation}
{ \Gamma_\text{L}= (-2.5 -1.7\beta + 0.1x)\frac{\Gamma_\text{0}}{\gamma_{\text{eff}}}},
\end{equation}
\begin{equation}
{\Gamma_{\text{c,hs,baro}}= 1.1\left( \frac{3}{2}-x\right) \frac{\Gamma_\text{0}}{\gamma_{\text{eff}}}},
\end{equation}
\begin{equation}
{ \Gamma_{\text{c,lin,baro}}= 0.7\left( \frac{3}{2}-x\right) \frac{\Gamma_\text{0}}{\gamma_{\text{eff}}}},
\end{equation}
\begin{equation}
{\Gamma_{\text{c,hs,ent}}= 7.9\xi\frac{\Gamma_\text{0}}{\gamma_{\text{eff}}^2}},
\end{equation}
and
\begin{equation}
 \Gamma_{\text{c,lin,ent}}= \left( 2.2 - \frac{1.4}{\gamma_{\text{eff}}}\right)\xi \frac{\Gamma_0}{\gamma_{\text{eff}}}
\end{equation}
where ${ \xi = \beta - (\gamma -1)x}$ is the negative of the entropy slope with $\gamma=$1.4 being the adiabatic index. The scaling torque ${ \Gamma_0=(q/h)^2\Sigma_{\text{gas}} r^4 \Omega_k^2}$ is defined at the location of the planet. The planet-star mass ratio is represented by q, h is the disk aspect ratio, ${ \Sigma_{gas}}$ is the surface density and  ${\Omega_k}$ is the planet's Keplerian orbital frequency.

Thermal and viscous diffusion effects  contribute differently to the different components of the coorbital torque. For example, the barotropic part of the coorbital torque is not affected by thermal diffusion while the entropy related part is affected by both thermal and viscous diffusion.  The parameter governing viscous saturation is defined by 
\begin{equation}
{p_{\rm \nu} = \frac{2}{3}\sqrt{\frac{r^2\Omega_k}{2\pi\nu}x_s^3}},
\end{equation}
where ${x_{\rm s}}$ is the non-dimensional half-width of the horseshoe region,
\begin{equation}
{ x_{\rm s}=\frac{1.1}{{\gamma_\text{eff}}^{1/4}}\sqrt{\frac{q}{h}}}.
\end{equation}

The effects of thermal saturation at the planet location are controlled by
\begin{equation}
{ p_{\chi} = \sqrt{\frac{r^2\Omega_k}{2\pi\chi}x_s^3}},
\end{equation}
where ${\rm \chi}$ is the thermal diffusion coefficient which reads as
\begin{equation}
{ \chi = \frac{16\gamma(\gamma -1)\sigma T^4}{3 \kappa \rho^2(hr)^2 \Omega_k^2}},
\end{equation}
where $\rho$ is the gas volume density, $\kappa$ is the opacity and $\sigma$ is the Stefan-Boltzmann constant. The other variables are defined before.

Finally we need to set
\begin{equation}
Q  = \frac{2 \chi}{3 h^3 r^2 \Omega_k} 
\end{equation}
to define the effective $\gamma$, used in Eq. 18, as
\begin{equation}
\gamma_\text{eff} =  \frac{2Q \gamma}{\gamma Q + \frac{1}{2}\sqrt{2 \sqrt{( \gamma^2Q^2+1)^2-16 Q^2(\gamma-1)}+2 \gamma^2 Q^2 - 2}}.
\end{equation}

The functions F, G, and K in Eq. 11 are given by

\begin{equation}
F(p) = \frac{1}{1+ \left( \frac{p}{1.3}\right)^2, }
\end{equation}
\begin{equation}
  G(p)=\begin{cases}
    \frac{16}{25}\left( \frac{45\pi}{8}\right)^{\frac{3}{4}}p^{\frac{3}{2}}, & \text{if $p< \sqrt{\frac{8}{45\pi}}$}\\
    1 - \frac{9}{25}\left( \frac{8}{45\pi}\right)^{\frac{4}{3}}p^{-\frac{8}{3}}, & \text{otherwise}.
  \end{cases},
\end{equation}
and
\begin{equation}
  K(p)=\begin{cases}
    \frac{16}{25}\left( \frac{45\pi}{28}\right)^{\frac{3}{4}}p^{\frac{3}{2}}, & \text{if $p< \sqrt{\frac{28}{45\pi}}$}\\
    1 - \frac{9}{25}\left( \frac{28}{45\pi}\right)^{\frac{4}{3}}p^{-\frac{8}{3}}, & \text{otherwise}.
  \end{cases}.
\end{equation}
Note that the $p$ takes the form of $p_{\nu}$ (Eq. 16) or $p_{\chi}$ (Eq. 18) as defined above.

Following \cite{papaloizoularwood00} and \cite{cresswellnelson08}  we define the planet's migration timescale as

\begin{equation}
{ t_m =- \frac{L}{\Gamma_{tot}}}.
\end{equation}
With this definition of $t_m$ the respective timescale for a planet on circular orbit to reach the star is $t_m/2$. In Eq. 26, the quantity L is a planet's orbital angular momentum and  ${\rm \Gamma_{tot}}$ is the type-I torque defined in Eq.~6 . 

To account for the effects of eccentricity and inclination damping we follow the classical formalism of \cite{papaloizoularwood00} and \cite{tanakaward04} modified by \cite{cresswellnelson06,cresswellnelson08}. Eccentricity and inclination damping timescales are given by  ${\rm t_e}$ and ${\rm t_i}$, respectively.  They are defined as



\begin{multline}
t_e = \frac{t_{wave}}{0.780} \left(1-0.14\left(\frac{e}{h}\right)^2 + 0.06\left(\frac{e}{h}\right)^3  \right. \\ \left. + 0.18\left(\frac{e}{h}\right)\left(\frac{i}{h}\right)^2\right),
\end{multline}
and
\begin{multline}
 t_i = \frac{t_{wave}}{0.544} \left(1-0.3\left(\frac{i}{h}\right)^2 + 0.24\left(\frac{i}{h}\right)^3 \right.  \\  \left.  + 0.14\left(\frac{e}{h}\right)^2\left(\frac{i}{h}\right)\right),
\end{multline}
where

\begin{equation}
 t_{wave} = \left(\frac{M_{\odot}}{m_p}\right)  \left(\frac{M_{\odot}}{\Sigma_{gas} a^2}\right)h^4 \Omega_k^{-1}
,\end{equation}
and ${\rm M_{\odot}}$, ${\rm a_p}$, ${\rm m_p}$, ${\rm i}$, ${\rm  and~e}$ are the solar mass and the embryo semimajor axis, mass, orbital inclination, and eccentricity, respectively. 

Using the previously defined timescales, the artificial accelerations to account for type-I migration, eccentricity and inclination damping included in the equations of motion of the planetary embryos in our simulations are namely
\begin{equation}
{\bold{a}_m = -\frac{\bold{v}}{t_m}},
\end{equation}

\begin{equation}
{\bold{a}_e = -2\frac{(\bold{v.r})\bold{r}}{r^2 t_e}},
\end{equation}
and
\begin{equation}
{\bold{a}_i = -\frac{v_z}{t_i}\bold{k},}
\end{equation}
where ${ \bold{k}}$ is the unit vector in the z-direction.  Eq. 30-32 are given in  \cite{papaloizoularwood00} and \cite{cresswellnelson06,cresswellnelson08}.

\subsection{A Migration map}
Combining our disk model and type-I migration recipe we can build a migration map showing the migration rate and direction as a function of a planet's semi-major axis and mass.

\begin{figure}
\centering
\includegraphics[scale=.5]{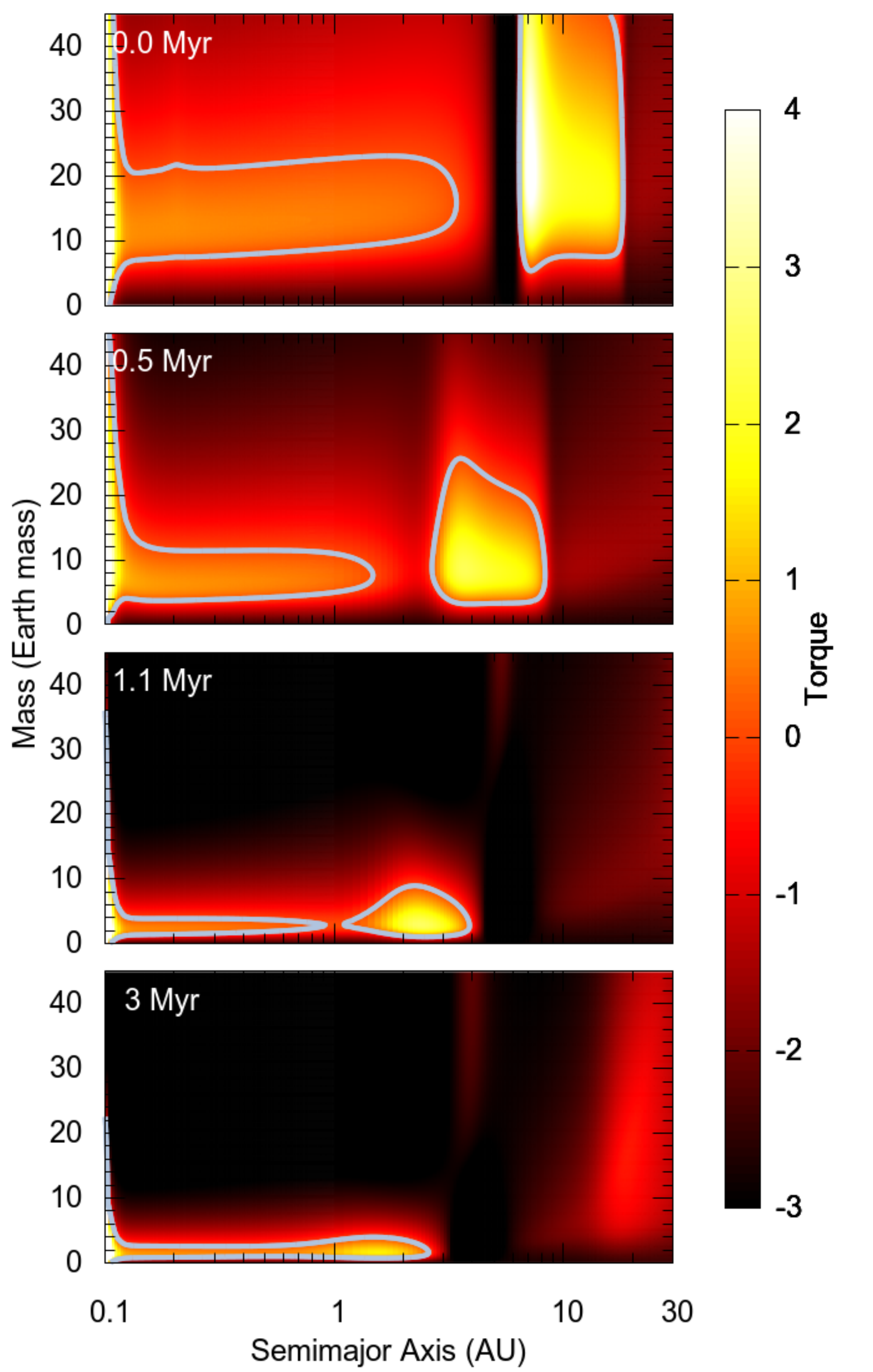}
\caption{Evolution of the migration map calculated in a disk with metallicity equal to 1\% and  $\alpha=0.0054$ (dimensionless viscosity). The gray lines in each panel  correspond to zero-torque locations and they delimit outward migration regions. At $t_{disc}$=0 yr  two different regions where outward migration is possible are shown. The most prominent one extends from about 5 to 20 AU for planets with masses from 10 to more than 40 Earth masses. As the disk evolves these regions moves inwards, shrink and eventually merge. The yellow region at about 0.1 AU corresponds to the planet trap set at the disk inner edge. }
    \label{fig:map}
\end{figure}

Figure~\ref{fig:map} presents an evolving migration map of our chosen disk.  It shows the direction and relative speed of migration of planets on circular and coplanar orbits at different locations within the disk. The direction of migration is represented by the color. A negative torque (reddish to black) implies inward migration while a positive one (orange to light yellow) represents outwards migration. The grey lines represent locations of zero torque.  The regions enclosed by the grey lines are areas of positive torque where planets migrate outward. The strong positive torque  at about 0.1 AU is a consequence of the imposed planet trap at the disk inner edge \citep{massetetal06}.

\section{Fiducial Model}

We performed 120 simulations of our fiducial model. Here we included the effects of type-I migration, eccentricity and inclination damping forces as described in Section 2.  In later sections we will present simulations that test the effects of turbulence.

\subsection{Initial conditions}

 Our simulations start from a population of 20-30 planetary embryos distributed beyond 5 AU.  This inner edge was chosen to approximately correspond to the location of the water snow line at the start of the simulations, for our chosen disk model~\citep{bitschetal15}. The embryos' initial masses are randomly and uniformly selected in the range from 0.1 to 4.5 Earth masses. The  total mass in planetary embryos in each simulation is about 60 Earth masses.  Adjacent planetary embryos were initially spaced by $\sim$5 mutual hill radii $R_{H,m}$\citep{kokuboida00}, where
\begin{equation}
R_{H,m} = \frac{a_i+a_j}{2}\left(\frac{M_i+M_j}{3 M_\odot}\right)^{1/3}.
\end{equation}
 In Eq. 33 $a_i$ and $a_j$ are the semi-major axes of the planetary embryos $i$ and $j$, respectively. Analogously, their masses are given by $M_i$ and $M_j$. In all simulations the time which planetary embryos start to evolve in the disk corresponds to ${t_{disk}}$=~0 yr. 

\subsection{Fiducial Model: dynamical evolution}

\begin{figure*}
	\includegraphics[width=1.05\linewidth]{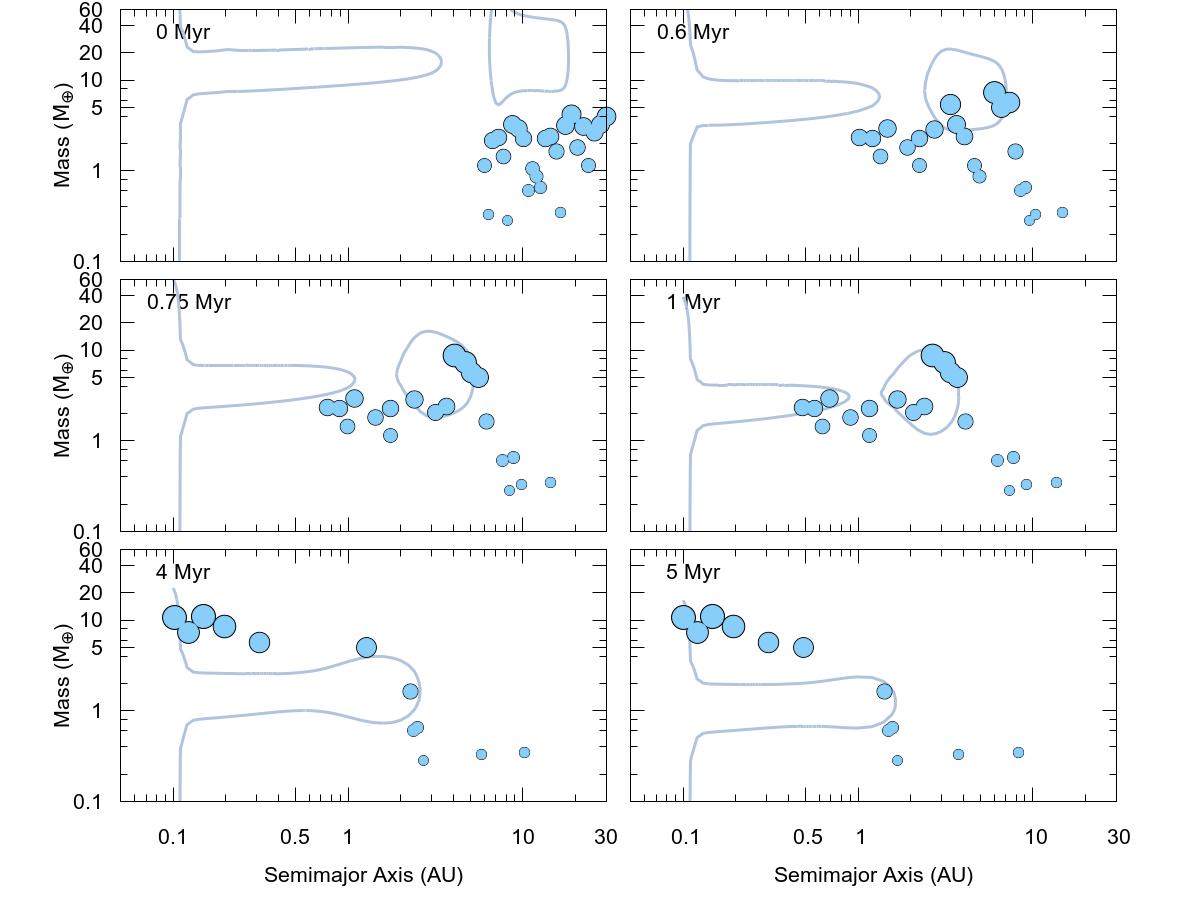}
    \caption{Evolution of a characteristic simulation of our fiducial set during the 5 Myr gas disk phase. Each blue circle corresponds to one embryo and the size of the point scales as $M_p^{1/3}$, where $M_p$ is the planet mass. The vertical axis shows the mass and the horizontal one shows the planet's semi-major axis. The gray dashed line delimits regions of outward migration and the disk inner edge. See also animated figure online. }
    \label{fig:mapplanets1}
\end{figure*}

\begin{figure*}
	\includegraphics[width=1.05\linewidth]{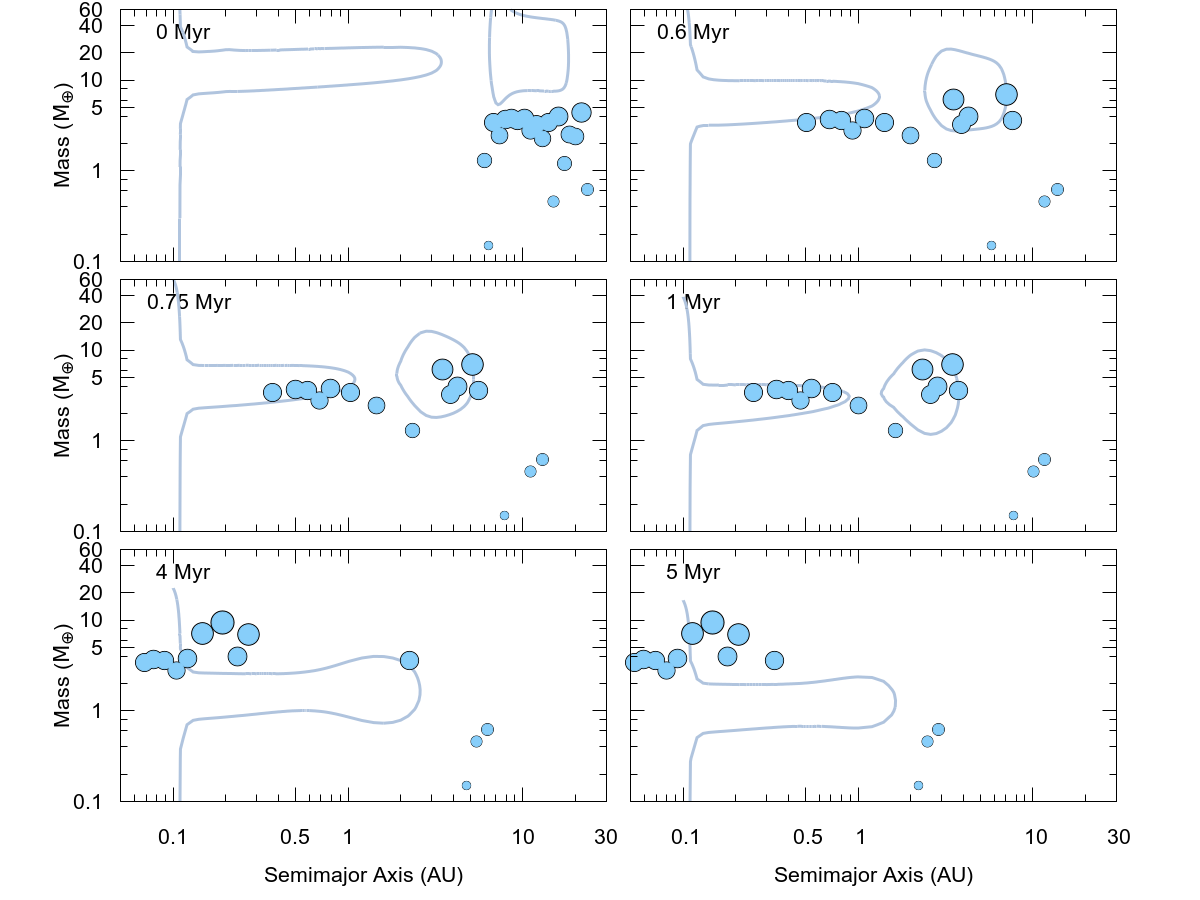}
    \caption{Another example of the dynamical evolution of the planetary embryos during the 5 Myr gas disk phase. The size of each blue circle scales as $M_p^{1/3}$, where $M_p$ is the planet mass. The vertical axis shows the mass and the horizontal one shows the planet's semi-major axis. The gray dashed line delimits regions of outward migration and the disk inner edge. See also animated figure online. }
    \label{fig:mapplanets2}
\end{figure*}

Figures \ref{fig:mapplanets1} and \ref{fig:mapplanets2} show the dynamical evolution of two characteristic simulations during the 5 Myr gas phase.  The gray lines in each panel delimit the outward migration regions shown in Figure \ref{fig:map}. The imposed planet trap at the disk inner edge at $\sim$ 0.1 AU is also evident. Embryos migrate inwards and converge to form resonant chains. Resonant planet-pairs often become unstable and collide.  As planetary embryos collide and grow (and the disk evolves) some enter the outward migration regions as it is possible to see in the panels corresponding to 0.6 and 0.75 Myr. Planets inside the outward migration region slowly migrate inwards, once the outward migration region also moves inward and shrinks \citep{lyraetal10}. Between 1 and 4 Myr the outward migration regions have evolved such that planets larger than 5 Earth masses migrate inward to the inner regions of the disk. This deposits planets in long resonant chains at the inner edge of the disk.  Small planetary embryos --typically smaller than 1 Earth mass migrate very slowly and stay beyond  1~AU. At the end of the gas disk-phase  planetary systems exhibits compact resonant configurations with 5-10 planets inside 0.5 AU.

\begin{figure*}
\centering
	\includegraphics[width=1.05\linewidth]{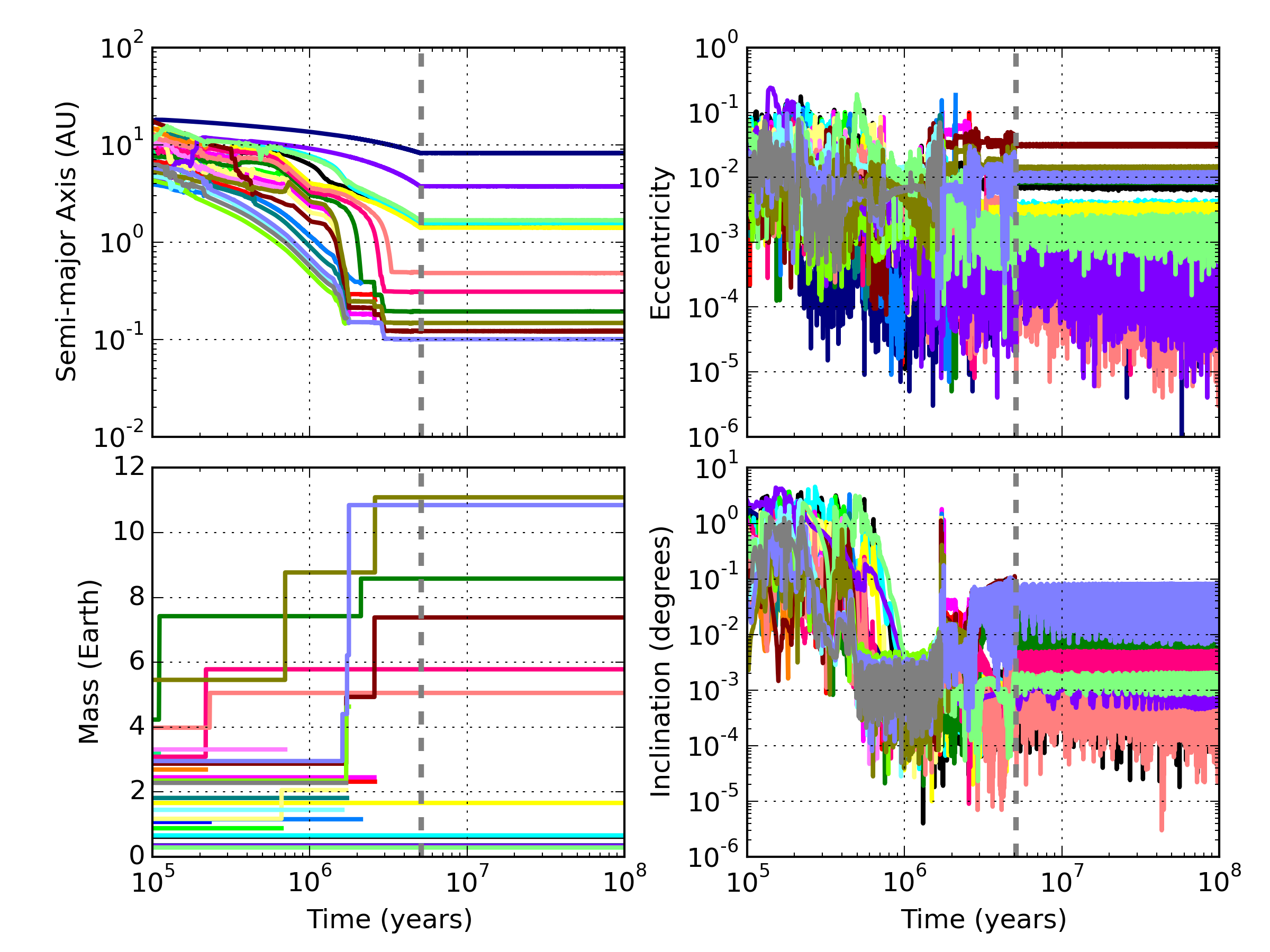}
    \caption{Dynamical evolution of planets in one simulation during and after the gas disk dispersal. The panels show the temporal evolution of planets' semi-major axis, eccentricity, mass and orbital inclination. The same line color is used to consistently represent a individual planet in all panels. The gas dissipates at 5.1 Myr and the system is numerically integrated up to 100 Myr. This planetary system is dynamically stable after the gas disk phase for at least 100 Myr. The gray dashed vertical line shows the time of the disk dissipation.}
    \label{fig:4panels_1}
\end{figure*}

\begin{figure*}
	\includegraphics[width=1.05\linewidth]{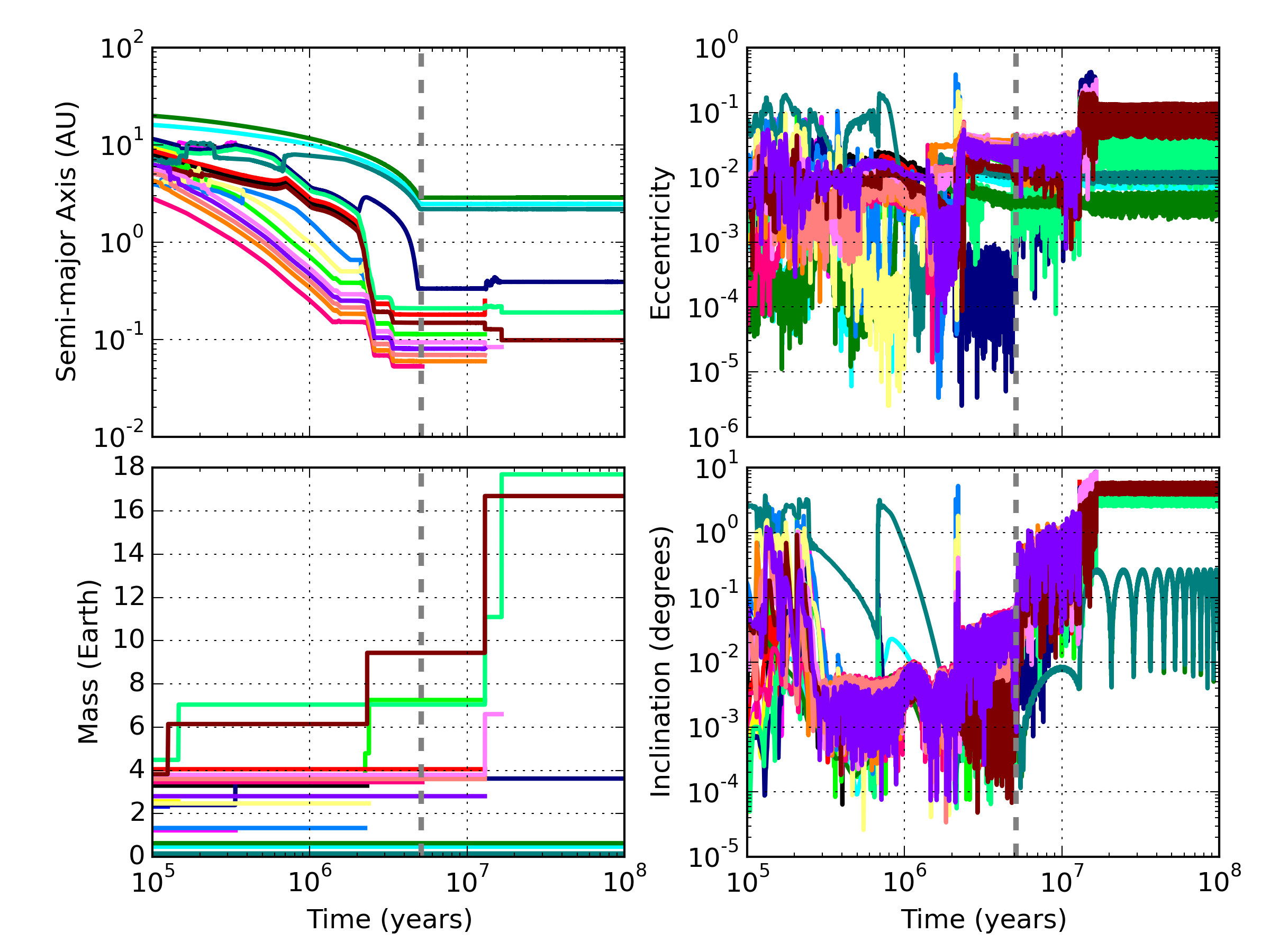}
    \caption{Dynamical evolution of planets in one simulation during and after the gas disk dispersal. The panels show the temporal evolution of planets' semi-major axis, eccentricity, mass and orbital inclination. The same line color is used to consistently represent a individual planet in all panels. The gas dissipates at 5.1 Myr and the system is numerically integrated up to 100 Myr. This planet system presents a phase of dynamical instability after the gas disk phase which lead to collisions and, consequently, to a planetary system dynamically less compact but relatively excited. The gray vertical line shows the time of the disk dissipation.}
    \label{fig:4panels_2}
\end{figure*}

Fig. \ref{fig:4panels_1} and \ref{fig:4panels_2} show the dynamical evolution during and after the gas-disk phase for the simulations shown in Fig.~\ref{fig:mapplanets1} and~\ref{fig:mapplanets2}. Recall that the gas lasts 5.1 Myr and the entire simulations last 100 Myr. The simulations from Figures~\ref{fig:4panels_1} and~\ref{fig:4panels_2} behaved in a similar fashion during the gas disk phase, but their later evolutionary paths are contrasting examples. The system from Figure~\ref{fig:4panels_1} remained stable after the gas disk dissipated, surviving in a long resonant chain containing 6 planets interior to 1 AU.  The system from Fig.~\ref{fig:4panels_2} underwent a series of instabilities that led to collisions and consequently, a planetary system that is dynamically less compact and more dynamically excited than the one from Fig.~\ref{fig:4panels_1}.  The final orbital eccentricities of planets shown in Figure~\ref{fig:4panels_1} are less than 0.05 and their orbital inclinations are smaller than 0.1 degree. The final  eccentricities of planets in Figure~\ref{fig:4panels_2} are as high as 0.1 and their inclinations are as high as 6 degrees. The most massive planets were somewhat larger in the unstable simulation as well ($\sim18 \mearth$ for the simulation from Fig.~\ref{fig:4panels_2} vs. $11 \mearth$ for the simulation from Fig.~\ref{fig:4panels_1}).

\section{Turbulent Disks}

Hydrodynamical instabilities in gaseous protoplanetary disk generate turbulence in the disk and transport angular momentum. Instabilities include the Rossby-wave instability \citep{lovelaceetal99}, the global baroclinic instability \citep{klahrbodenheimer03}, the Kelvin–Helmholtz instability generated during dust vertical sedimentation towards the disk midplane \citep{Johansenetal06}, and the vertical shear instability~\citep{nelsonetal13,stollkley14}. Another potentially important source for the observed gas accretion rate on young stars is turbulence driven by the magnetorotational instability \citep[MRI]{balbushawley98}. In a sufficiently ionized and magnetized disk the MRI generates magnetohydrodynamic turbulence which leads to outward angular momentum transport \citep{brandenburgetal95,armitage98}. MRI turbulence produces large-scale, axisymetric and long-lived density and pressure perturbations in the disk \citep{hawleyetal96}. Here, we perform simulations testing the role of this kind of turbulence for the formation of close-in super-Earths by inward migration. We assume that turbulence operates at levels consistent with estimates from  3D magnetohydrodynamic simulations ~\citep[e.g.][]{laughlinetal04,baruteaulin10}.  However, one should also note that recent studies have proposed that a number of  non-ideal effects can suppress magneto-rotational turbulence in a large region of the disk \citep{turneretal14}.

The effects of stochastic forcing on planet migration has been studied by several authors \citep{nelsonpapaloizou03,papaloizounelson03,winteretal03,laughlinetal04,nelson05,ogiharaetal07,adamsetal08,reinpapaloizou09,leconaetetal09,nelsoncressel10,
baruteaulin10,pierensetal11,ketchumetal11,hornetal12,pierensetal12,rein12}. However, the effects of turbulence for the origins of close-in super-Earths remains to be carefully addressed.

To model the turbulent motion of the gas, which essentially corresponds to density fluctuations in the gaseous disk, we use the model by \citep{laughlinetal04} as modified by \cite{ogiharaetal07} and \cite{baruteaulin10}. The specific turbulent force is given by

\begin{equation}
{\bf F }=  -\Gamma \nabla {\bf \Phi}
\end{equation}  
where
\begin{equation}
\Gamma = \frac{10^3\Sigma_{gas} r^2}{\pi^2 M_\bigstar}.
\end{equation}
Note that our $\Gamma$ is larger than that in \cite{laughlinetal04} by  a factor of $\sim$20 as found by \citep{baruteaulin10}. The potential induced by the turbulent perturbations consists of the sum of N independent, scaled wave-like modes as 
\begin{equation}
\Phi = \gamma r^2 \Omega^2 \sum_{i=1}^{N}\Lambda_{c,m}
\end{equation}
where a single oscillation mode is defined as
\begin{equation}
\Lambda_{c,m} = \xi e^{-\frac{(r-r_c)^2}{\sigma^2}}cos(m\theta - \phi_c + \Omega_c \tilde{t} )sin(\pi\frac{\tilde{t}}{\Delta t}).
\end{equation}
In Eq. 35, according to \cite{baruteaulin10} the strength of the potential is given by
\begin{equation}
\gamma =0.085h\sqrt{\alpha}.
\end{equation} 
Also, as noted in Eq. 36, a specific mode is determined by the azimuthal wavenumber $m$, which we randomly sort with a logarithmic distribution between 1 and 96, the center of its initial radial  location $r_c$, and azimuthal phase $\phi_c$. Only modes with wavenumber $m$ smaller than 6 are considered \citep{ogiharaetal07}. To sample $r_c$ we use a log-normal distribution to select values between ${\rm r_{in}}$ and ${\rm r_{out}}$. In our simulations ${\rm r_{in} = 0.1AU}$ and ${\rm r_{out} = 25AU}$. The outer edge of the turbulent region (${\rm r_{out}}$) roughly corresponds to the initial location of the outermost planetary embryo of the system. Thus, our planet formation/migration region is significantly wider than that in \cite{baruteaulin10} and \cite{ogiharaetal07}. Because of this we assume the existence of a larger number of wave-like modes N in the disk than most of previous studies. We set ${\rm N=125}$, similar to \cite{hornetal12}. $\phi_c$ is sorted with a uniform distribution between 0 and 2$\pi$. The dimensionless parameter $\xi$ is sorted according to a Gaussian distribution with a unitary standard deviation and mean-value zero. Each mode has an radial extent of $\sigma = \pi r_c /4m$. The planet coordinates are represented by the radial distance $r$ and its azimuthal coordinate $\theta$. $\Omega_c$ is the Keplerian frequency calculated at $r_c$.  Still, turbulent fluctuations appear and disappear in the disk. To account for this phenomenon in Eq. 35, the wave mode lifetime is defined by $\Delta t=0.2\pi r_c/m c_s$ \citep{baruteaulin10}, where $c_s$ is the local sound speed. Thus, a given mode $m$  may only exist from its birth time $t=t_0$ to $\tilde{t} = t - t_0~=~ \Delta t$. If $\tilde{t}> \Delta t$ then  a new mode is created to take the  extinguished one's place.

Finally, according to \cite{ogiharaetal07} the radial, azimuthal and vertical components of the  artificial force to account for the effects of turbulence may be written as 

\begin{equation}
F_{turb,r} = \gamma \Gamma r \Omega^2  \sum_{i=1}^{N} \left( 1 + \frac{2r(r-r_c)}{\sigma^2} \right) \Lambda_{c,m},
\end{equation}

\begin{equation}
F_{turb,\theta} = \gamma \Gamma r \Omega^2  \sum_{i=1}^{N}  \Lambda_{c,m},
\end{equation}
and
\begin{equation}
F_{turb,z} = 0~.
\end{equation}

For simplicity, in our model we assume that the there is no feedback from the stochastic density/pressure fluctuations in the disk on our disk structure model. Basically, we assume the same underlying disk model than that used for our fiducial simulations.

\subsection{Simulations}

\begin{figure*}
	\includegraphics[width=1.05\linewidth]{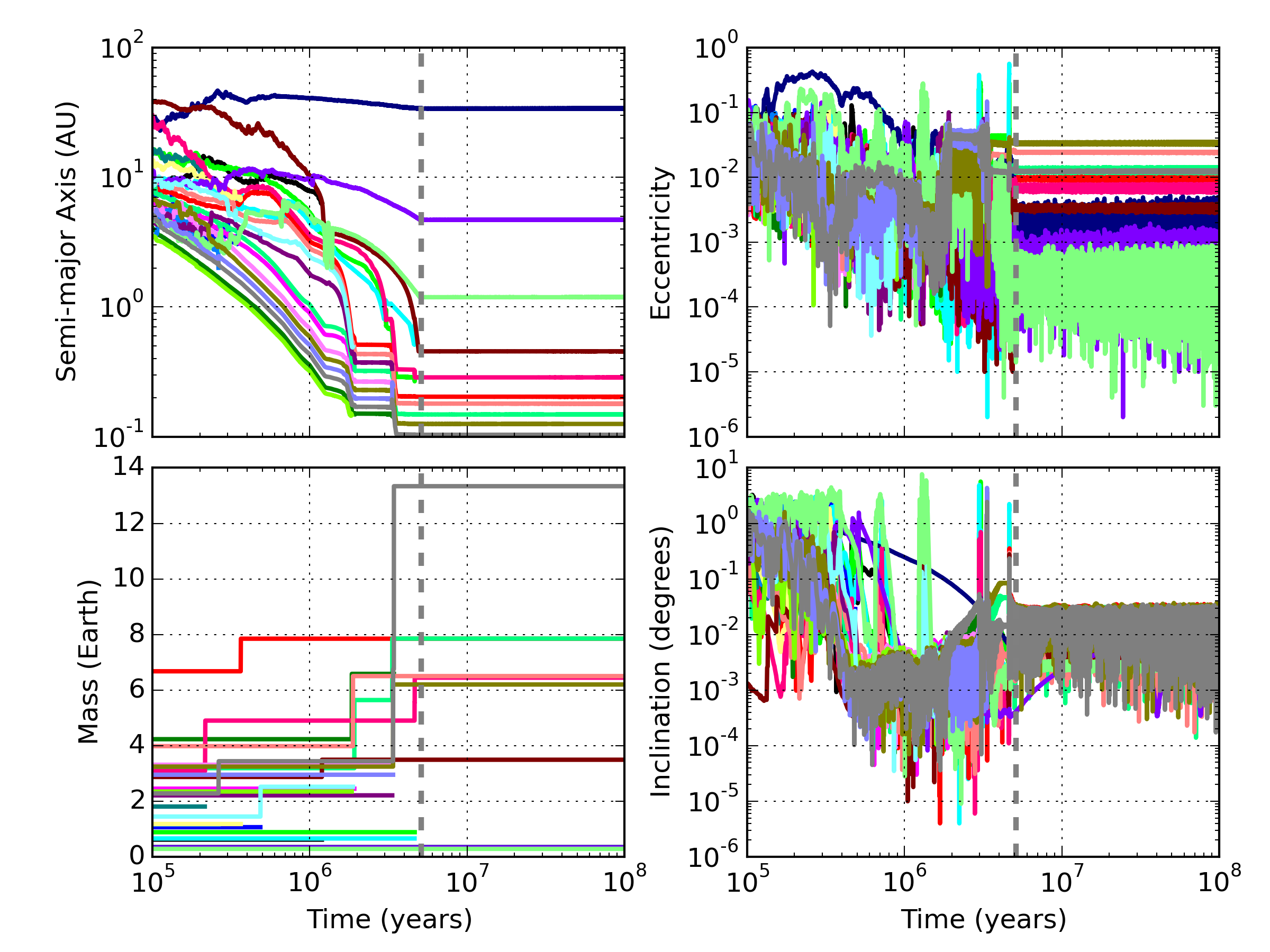}
    \caption{Dynamical evolution of planets in one simulation including the turbulent effects during and after the gas disk dispersal.  The panels show the temporal evolution of planets' semi-major axis, eccentricity, mass and orbital inclination. The same line color is used to consistently represent each  planet in all panels. The gas dissipates at 5.1 Myr and the system is numerically integrated up to 100 Myr. This planetary system is dynamically stable after the gas disk phase for at least 100 Myr. The gray vertical line shows the time of the disk dissipation.}
    \label{fig:4panels_3}
\end{figure*}

We performed 120 simulations with initial conditions identical to those from our 120 fiducial simulations but with the effects of turbulence. To illustrate the dynamical evolution of the planets in one characteristic simulation including turbulent effects we used the same initial distribution of planetary embryos as in Figure~\ref{fig:4panels_1}. The result of this simulation is shown in Figure~\ref{fig:4panels_3}. 

Figure~\ref{fig:4panels_3} shows that turbulence is important in the outermost parts of the disk, visible by the ``random walks'' in semi-major axis of planets beyond $\sim$5~AU. Turbulence also tends to increase the orbital eccentricity of planets in this region \citep{ogiharaetal07,rein12}. However, as planets migrate inward the effects of turbulence weakens. When planets reach the inner edge of the disk the turbulence is essentially negligible. It is easy to understand this result by inspecting our turbulent model. In our model, the extent of the fluctuation mode scales with $r_c$ (radial coordinate of the mode). In addition, assuming an aspect ratio of 3\% for the inner regions of the gaseous disk and a given $m$, the mode lifetime scales as $\Delta t \sim r_c^\frac{3}{2}$. Thus, modes in the inner regions of the disk (e.g. inside 1 AU) are also short-lived and this implies that they have modest to negligible contributions in disturbing the orbit of those planets (see also discussion  on the effects of using large wavenumbers in \cite{ogiharaetal07}).  Modes generated farther out, on the other hand, may be excessively far from planets already reaching the inner regions of the disk, to be able to strongly interact with them. In addition, it is also important to recall that the turbulence strength scales with the local gas surface density. At one given location,  as the disk dissipates its effects becomes relatively weaker. There is thus no significant difference between the results of our fiducial model and the results from that including effects of turbulence. Section 7 will discuss in some details why our results are different from others \citep{adamsetal08,rein12} in this respect.

 In the next two Sections we perform a careful statistical analysis of the results of our fiducial and turbulent models. We will interpret the origins of  their eventual differences and compare these results with observations and other works in the literature.

\section{Simulation outcomes}

Our simulations follow a bifurcated evolutionary path.  During the gas disk phase, planetary embryos grow and migrate inward to the inner edge of the disk.  The planets settle into long chains of mean motion resonances. It is important to recall that the resonant chains  are established very early, so a disk with a reduced lifetime would not help in avoiding these configurations. Resonant chains are typically established in less than 1.5~Myr (e.g. Figures~\ref{fig:4panels_1}-\ref{fig:4panels_3}).  After the gas disk dissipates, a large fraction of super-Earth systems undergo a dynamical instability.  The planets' orbits cross, leading to a phase of collisions that destroys the resonant chain.  However, a fraction of resonant chains remain stable and never undergo a late phase of collisions.  

We now analyze more in detail the evolution and outcome of simulations.  We present both the fiducial and turbulent sets of simulations.

\begin{figure}
  \includegraphics[width=.99\linewidth]{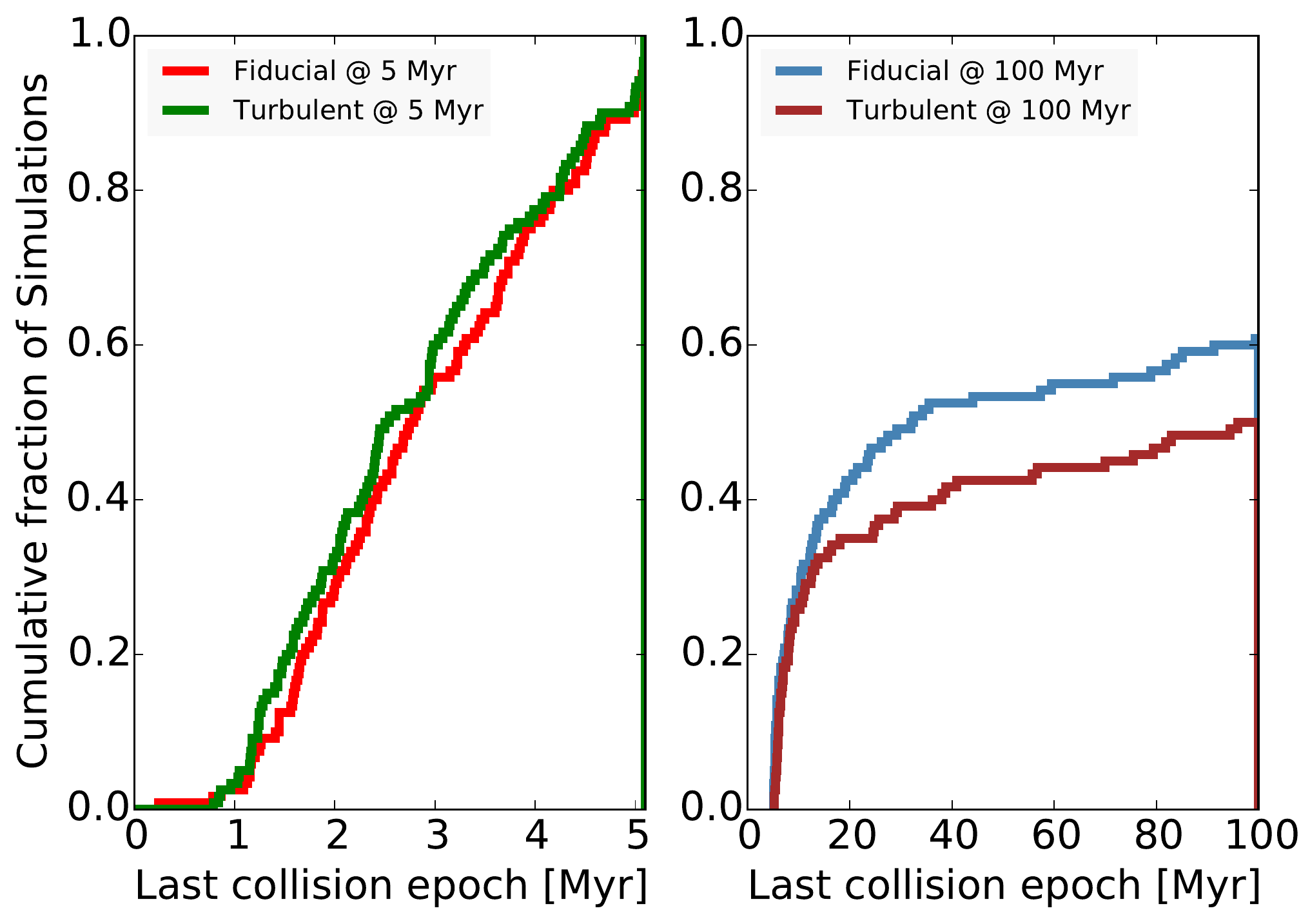}
    \caption{Cumulative distribution of the time of the last collision in our simulations. Cumulative distributions are separately calculated for
    the gas disk phase and post-gas disk dissipation phase. {\bf Left:} Computed from collisions happening between 0 and 5~Myr. {\bf Right:} Computed from collisions happening from 5 to 100~Myr.}
    \label{fig:last_coll}
\end{figure}

Figure~\ref{fig:last_coll} shows the cumulative distribution of the time of the last collision in our simulations during two different epochs: from 0 to 5~Myr and from 5 to 100~Myr. During the gas disk phase most systems have their last collisions after 1~Myr (left panel). This is expected since this corresponds to when most planets are approaching the disk inner edge (see for example Figs~\ref{fig:4panels_2} and~\ref{fig:4panels_3}), and have already reached relatively more compact configurations. This generates dynamical instabilities and collisions. Figure~\ref{fig:last_coll} also shows that during the gas disk phase the cumulative distributions of our fiducial and turbulent models are  similar.  The cumulative distributions of the last collision epoch grows broadly at constant rate from 1 to 4.5~Myr, both in our fiducial and turbulent model simulations. Thus, from 1.5 to 5~Myr there is no preferential time for last collision take place and every forming planetary system exhibit at least a few collisions during the gas disk phase. Because the disk is still present, the eccentricities and inclinations are damped again after each collision and the resonant chain is recovered due to the effect of residual migration.

After gas dispersal most late collisions occur during the first 20~Myr (right panel of Fig.~\ref{fig:last_coll}). Dynamical instabilities events tend to start as soon as the gas goes away (or when the gas becomes sufficiently rarefied). After about 20~Myr the rate of collisions start to drop. Also, the cumulative distributions do not reach 1. This indicates that not all planetary systems underwent instabilities after the gas disk phase (or did not experience any collision). About 60\% of the fiducial simulations and 50\% of the turbulent simulations experienced dynamical instabilities and at least one collision after the gas disk phase. Comparatively,  only 20\% of the in-situ formation simulations of \citep{ogiharaetal15a} were dynamically unstable after gas dispersal. This is  probably a consequence of the very small final number of planets in their planetary systems. We also recognize that the fraction of simulations  presenting dynamical instabilities may increase if our simulations were integrated for longer than 100~Myr. We are limited in the sense of extending the integration time of these simulations to  Gyr timescales because of the very small timestep necessary to resolve the orbits of planets that reach the inner orbit of the disk and the consequent long CPU time demanded.  However, we do not expect a linear growth with time of the number of unstable systems.

\begin{figure*}
	\includegraphics[width=.9\linewidth]{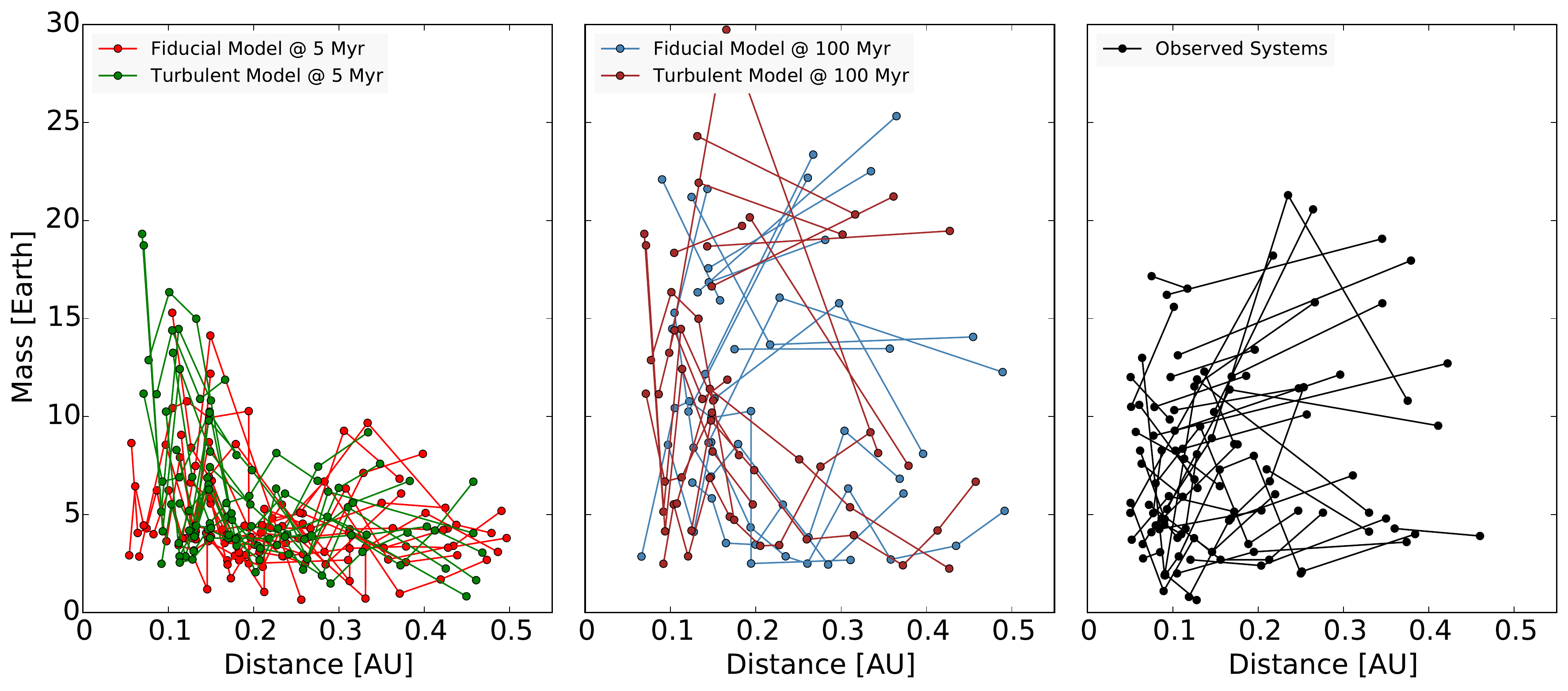}
    \caption{Representative sample of planetary systems produced in our fiducial and turbulent models at 5 and at 100~Myr. The vertical axis shows the mass and the horizontal one represents the planets' semi-major axis. {\bf Left:} Planetary systems at 5 Myr. {\bf Middle}: Planetary systems at 100 Myr. For each model we randomly selected 20 planetary systems which are plotted together. {\bf Right}: Selected observed planetary systems. The line connects planets in the same system. }
    \label{fig:planetarysystems}
\end{figure*}

Figure~\ref{fig:planetarysystems} presents representative simulated systems at 5 and 100~Myr.  For reference, we also show  selected observed planetary systems. Lines connect planets belonging to the same planetary system. More massive planets tend to park preferentially at the inner edge of the disk (Figure~\ref{fig:planetarysystems}, left panel). This is  because more massive planets simply migrate faster than small ones and just scatter outward  or collide with small ones during their radial excursion to the innermost regions of the disk \citep{izidoroetal14b}. However, we stress that there is no dramatic mass-ranking in our model, in contrast with systems of super-Earths produced by in-situ accretion~\citep{ogiharaetal15a}. This is a strong argument that favours the migration model over the in-situ model.  After the gas dissipates, instabilities largely erase the mass-ranking (Figure~\ref{fig:planetarysystems} -  middle panel).

\subsection{Architecture of planetary systems before gas dispersal}

\begin{figure}
     \includegraphics[width=.9\linewidth]{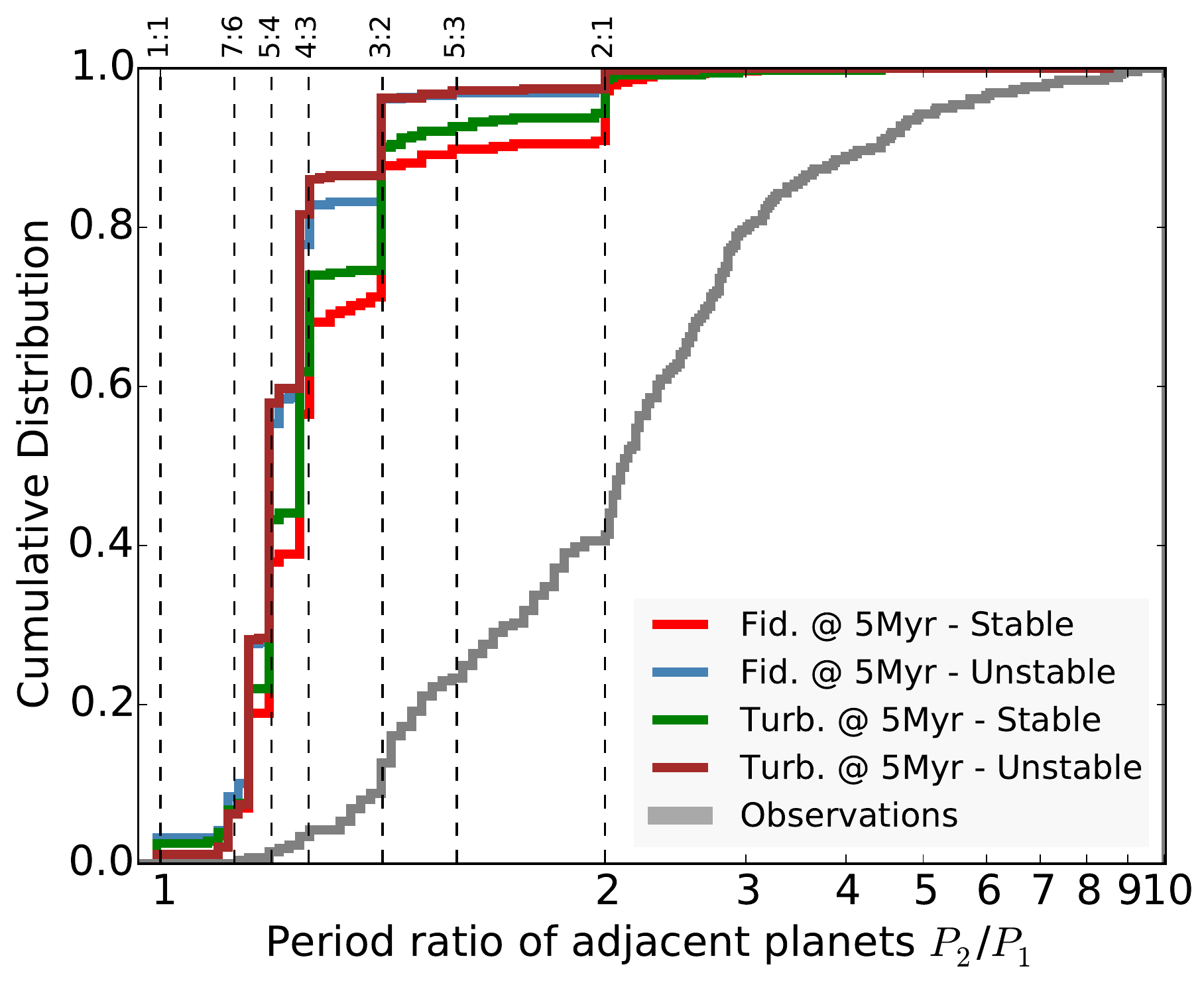}
     \includegraphics[width=.9\linewidth]{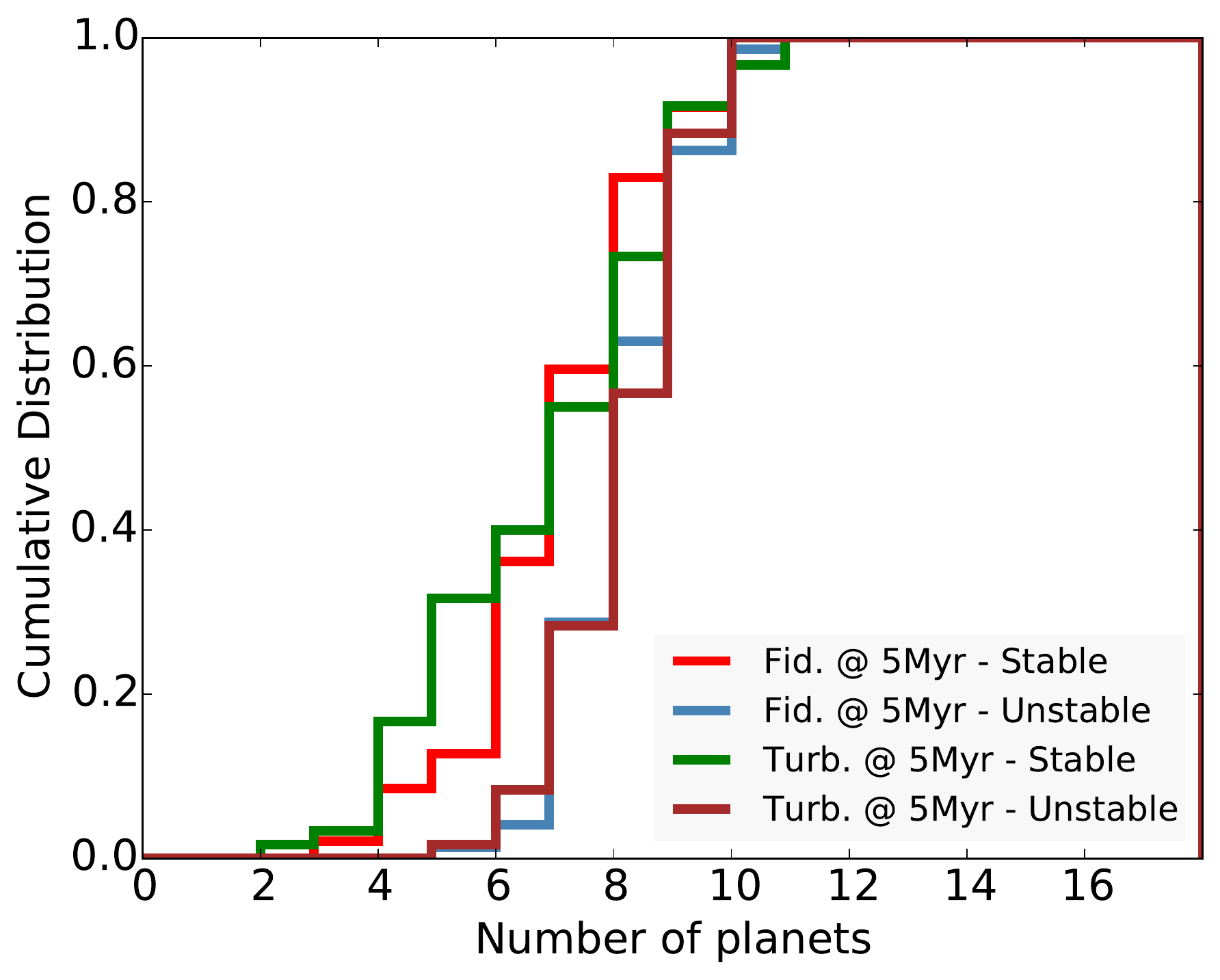}
     \includegraphics[width=.9\linewidth]{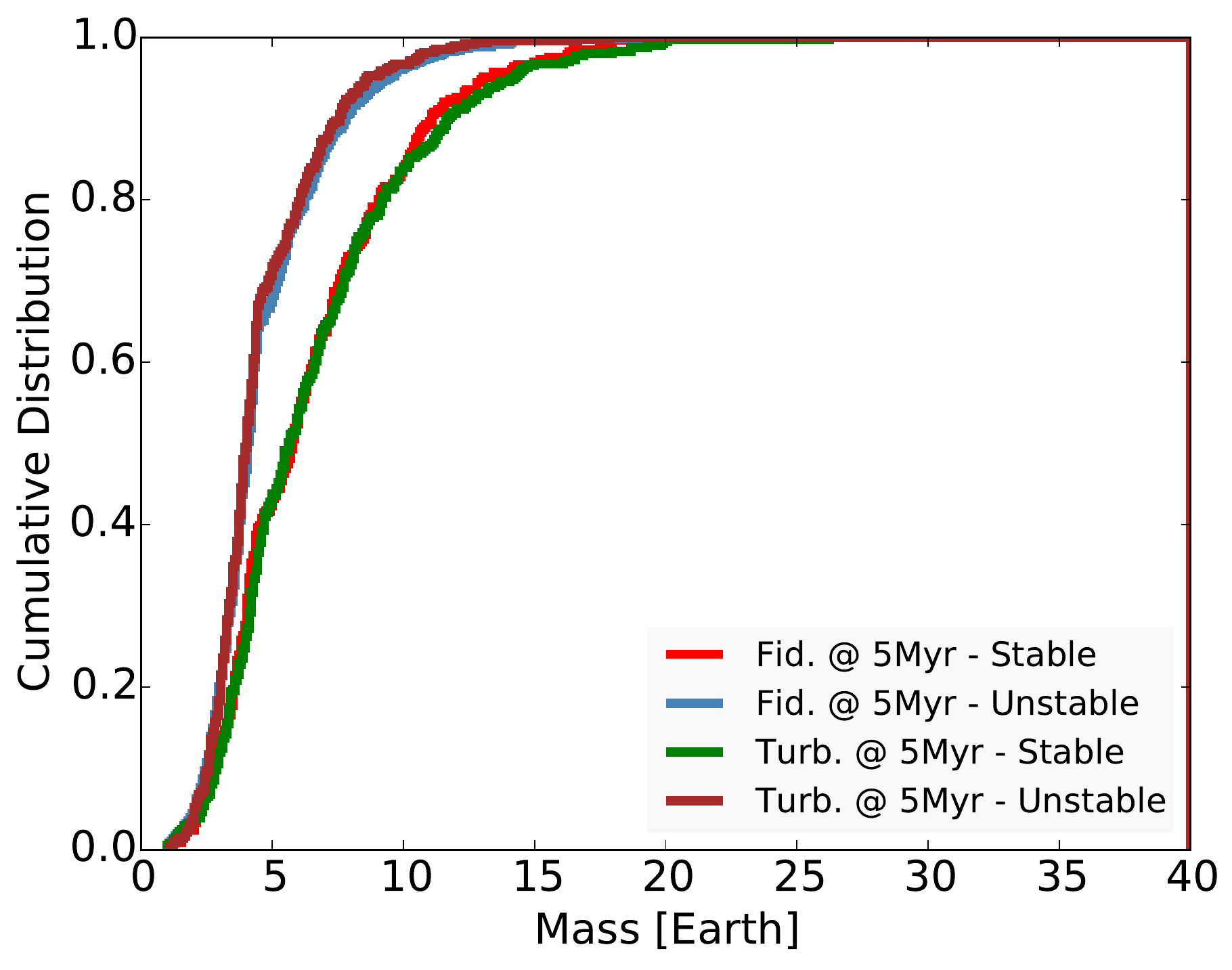}
   \caption{Cumulative distributions of period ratio of adjacent planets ({\bf top}), number of planets in the resonant chain ({\bf middle}), and masses of planets ({\bf bottom}) at 5~Myr. Only planets inside 0.5~AU are considered. The gray line in the top-panel shows the period ratio distribution of adjacent planet pairs in Kepler systems.} 
    \label{fig:unstable_vs_stable}
\end{figure}

Our simulations can be separated into two groups: those that underwent a late dynamical instability after the gas disk dissipated, and those that remained dynamically stable. We often refer to these as {\it stable} and {\it unstable} systems. It is important to keep in mind that the unstable systems started out as resonant chains. Thus, it is worth first to investigate how unstable planetary systems compare to stables ones  before the gas dispersal, namely at 5~Myr. The question we want to address is the following: Is there any systematic difference in the architecture of unstable and stable systems before gas dispersal? To answer this question we now  separate the simulations of our fiducial and turbulent models in stable and unstable groups. Thus, we are left with four sets of simulations which we naturally nominate: fiducial-stable, fiducial-unstable, turbulent-stable and turbulent-unstable. The results of our analysis for each of these groups are shown in Figure~\ref{fig:unstable_vs_stable}.

We first focus on the results of our fiducial model. Observing the period ratio distribution of adjacent planet pairs in Figure~\ref{fig:unstable_vs_stable} (top-plot) it is clear that at 5~Myr planetary systems are found in compact first order mean motion resonances (seen as the vertical lines in the plot). Yet, a more clinical analysis reveals that unstable planetary systems are  slightly more dynamically compact than stable ones (at 5~Myr). Figure~\ref{fig:unstable_vs_stable} (middle-plot) also shows that at 5~Myr planetary systems have typically multiple planets. Interestingly, the typical number of planets of stable chains is smaller than the number of planets of unstable ones. Figure~\ref{fig:unstable_vs_stable} also show how the masses of planets in stable and unstable system compare to each other at 5~Myr. The mass cumulative distributions of Figure~\ref{fig:unstable_vs_stable} (bottom-plot) show that planets in stable systems are predominantly more massive than those in unstable simulations. Finally, note that these trends are robust since they are observed also in the turbulent sets of simulations.

The results of Figure~\ref{fig:unstable_vs_stable} are quite intuitive since we expect more compact systems to be more prone to exhibit dynamical instabilities than  more spread out ones. However, it is important to recall that although there are some quantitative differences,  planetary systems shown in Figure~\ref{fig:unstable_vs_stable} are qualitatively similar in some key aspects. The most notable one is that, before gas dispersal, planets pairs are essentially found in compact resonant configurations. However, this is about to change for unstable systems. While stable systems configurations remains essentially unchanged during 100~Myr, in the next Section we show how dynamical instabilities sculpt unstable planetary systems.

\subsection{The importance of dynamical instabilities}

 In this section we analyse the architecture of unstable planetary systems at two epochs: before gas dispersal (at 5Myr) or, equivalently, ``before instability'' and at 100~Myr (``after instability''). Our main goal is to compare how dynamical instabilities shape the architecture of planetary systems. Note that the dynamical architectures of stable systems are essentially identical at 5 and 100~Myr. 
 
\begin{figure}
      \includegraphics[width=.9\linewidth]{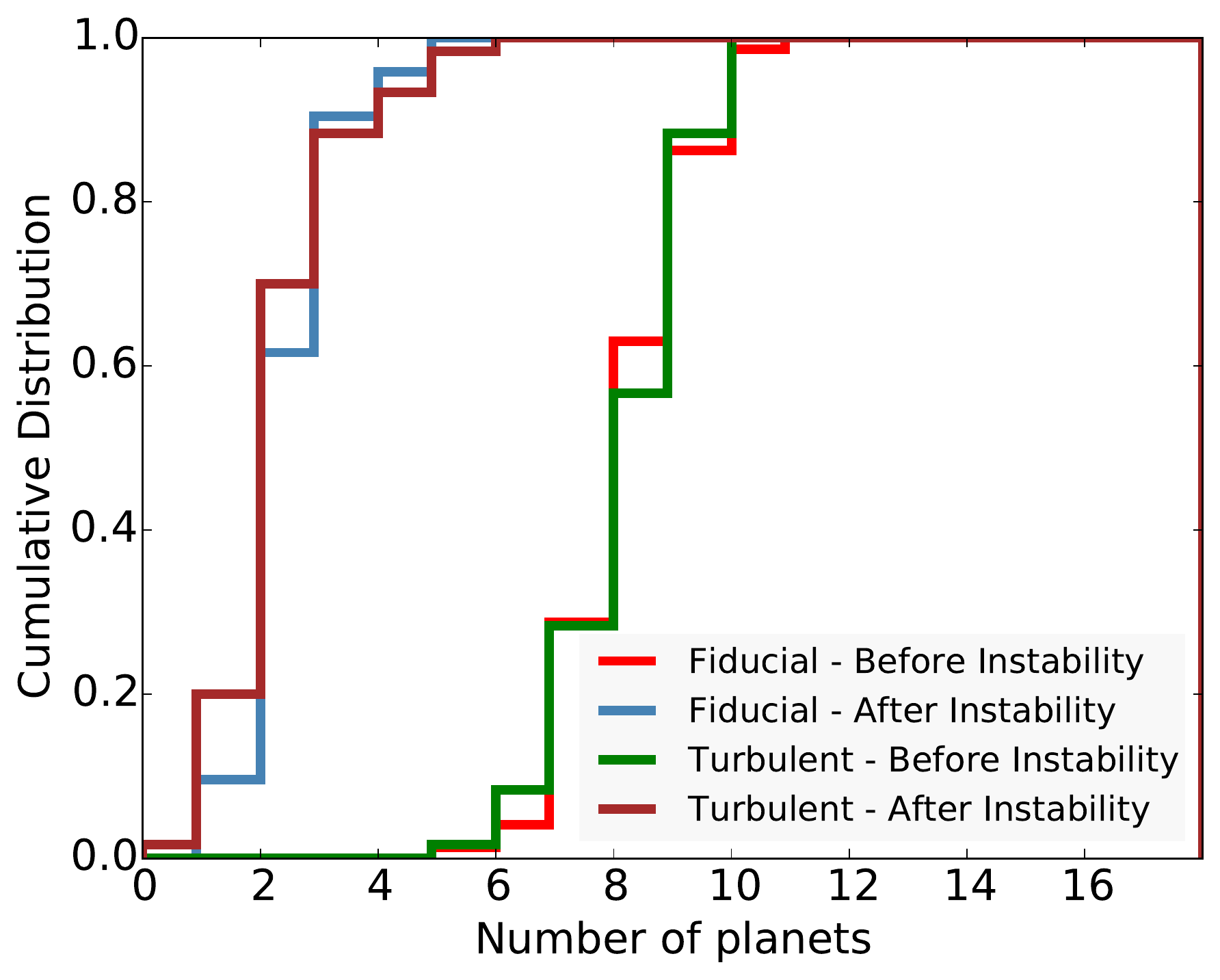}
    \caption{Cumulative distribution of the total number planets inside 0.5 AU produced in our fiducial and turbulent model. The cumulative distributions are shown at 5 and at 100 Myr. Only unstable planetary systems are considered here.}
    \label{fig:N_chain}
\end{figure}

Figure~\ref{fig:N_chain} shows the number of planets in unstable planetary systems before and after gas dispersal, or equivalently before and after the dynamical instability phase.  Given how compact resonant chains are, they typically contain 6-10 planets inside 0.5 AU.  However, given that instabilities spread out the system and reduce the number of planets (due to collisions), the typical unstable system contains 2-5 planets inside 0.5 AU.  

There is no glaring difference between the populations of resonant chains in the fiducial and turbulent simulations.  Yet roughly 10\% more fiducial simulations were unstable at later times (Figure \ref{fig:last_coll}). This difference probably comes from a subtle difference in some characteristic of the resonant chains (e.g.  number of planets in the chain, commensurabilities and masses of planet-pairs~\citep{matsumotoetal12}, amplitude of libration of resonant angles~\citep{adamsetal08}, etc). Perhaps part of this different is also due to small number statistics.

\begin{figure}
    \includegraphics[width=.9\linewidth]{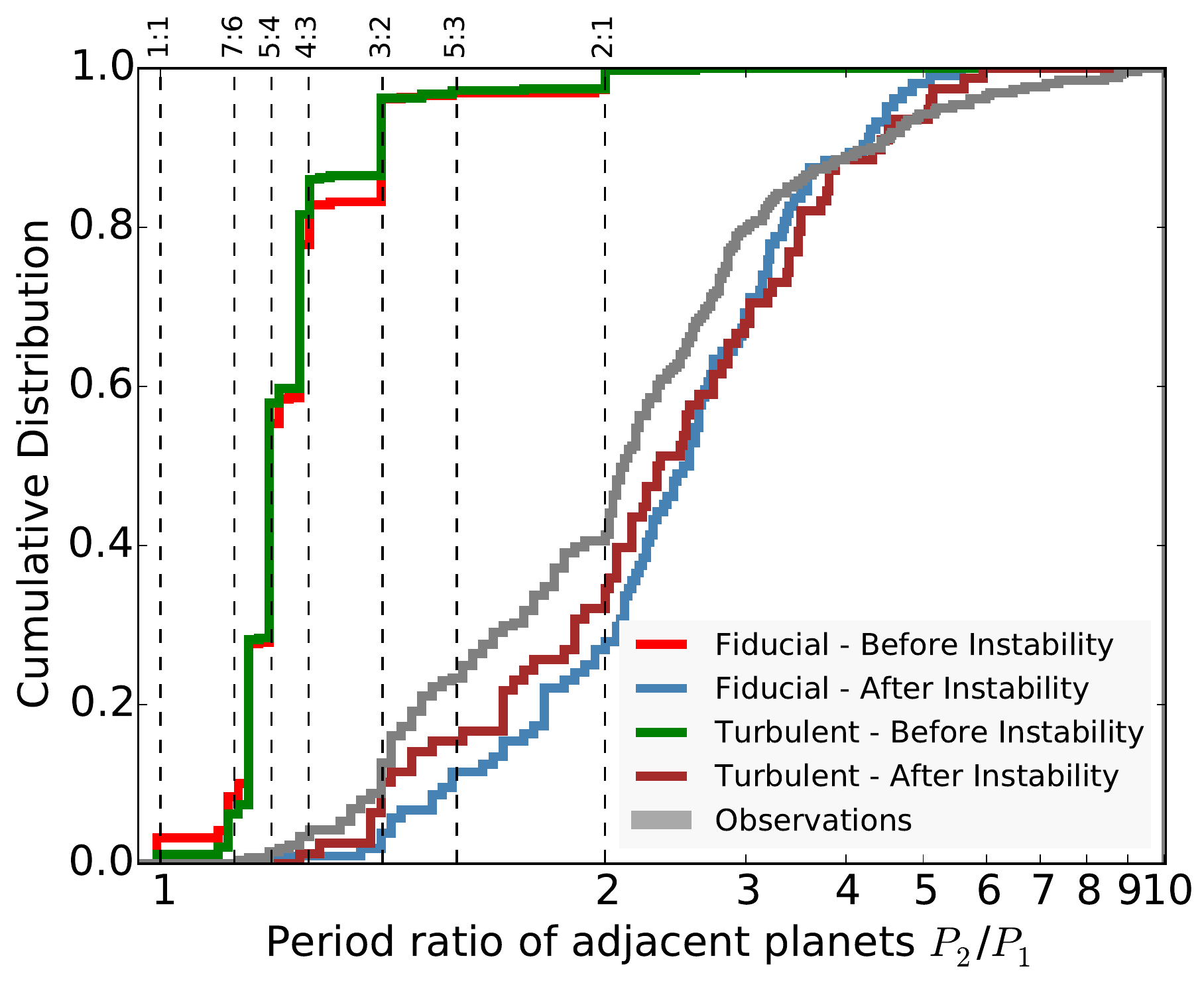}
    \caption{Cumulative period ratio distributions of adjacent planets produced in our fiducial and turbulent model for dynamically unstable systems, before and after the dynamical instability. The grey line corresponds to the observed period ratio of adjacent planets in the Kepler data.}
    \label{fig:periodratio_unstable}
\end{figure}

Figure~\ref{fig:periodratio_unstable} shows the period ratio distribution in systems that underwent dynamical instabilities after the gas dispersed. We also add the sample of planet candidates from the Kepler mission~\citep{boruckietal11,batalhaetal13,roweetal14}. To better compare our simulations with observations we applied a simple filter to the observations and our simulations. We only included planets with orbital radii smaller than 0.5 AU.  In the Kepler data we included planets with orbital period shorter than 130 days and radii smaller than 4 Earth radii. The motivation for choosing these cutoffs comes from the completeness of Kepler observations \citep{silburtetal15}. We advance to the reader that to account for the effects of inclination distribution we will perform simulated observations in Section 6.

At the end of the gas disk phase at 5 Myr, planets are universally found in chains of mean motion resonances, seen as the vertical jumps in Fig.~\ref{fig:periodratio_unstable}. Resonant chains are far more compact than the observed systems. These are  the ``before instability'' systems (see also the stable systems in Figure~\ref{fig:unstable_vs_stable}).  There is little difference between the fiducial and turbulent simulations. Instabilities break the resonant chains created during the gas disk phase, promote collisions and scattering events that reduce the number of planets in the system (Figure~\ref{fig:N_chain}). This tends to produce planetary systems with planets far more apart from each other, and also on orbits with higher eccentricities and orbital inclinations (compare Figures~\ref{fig:4panels_1} and \ref{fig:4panels_2}).  This same trend is observed in the turbulent simulations. After dynamical instabilities, our planetary systems are less compact than the observed systems, at least for period ratios smaller than 3\footnote{The tail of the period ratio distribution probably suffer from selection bias effects.}. This suggest that dynamical instabilities play a crucial role in sculpting system of super-Earths.~\citep[see also][who proposed that the Kepler systems were sculpted by instabilities but without starting from resonant chains]{puwu15}

\begin{figure*}
     \includegraphics[width=.45\linewidth]{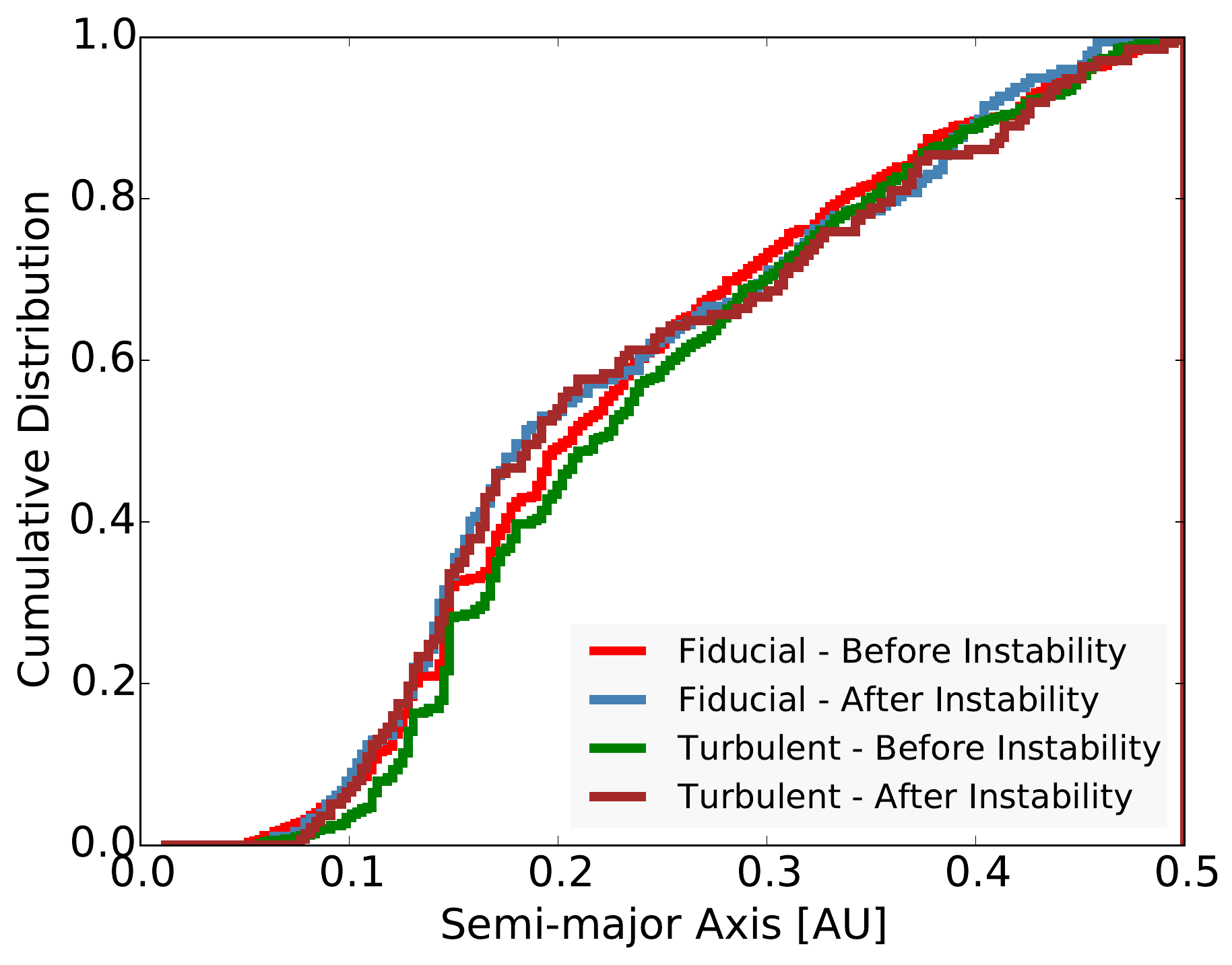}
     \includegraphics[width=.45\linewidth]{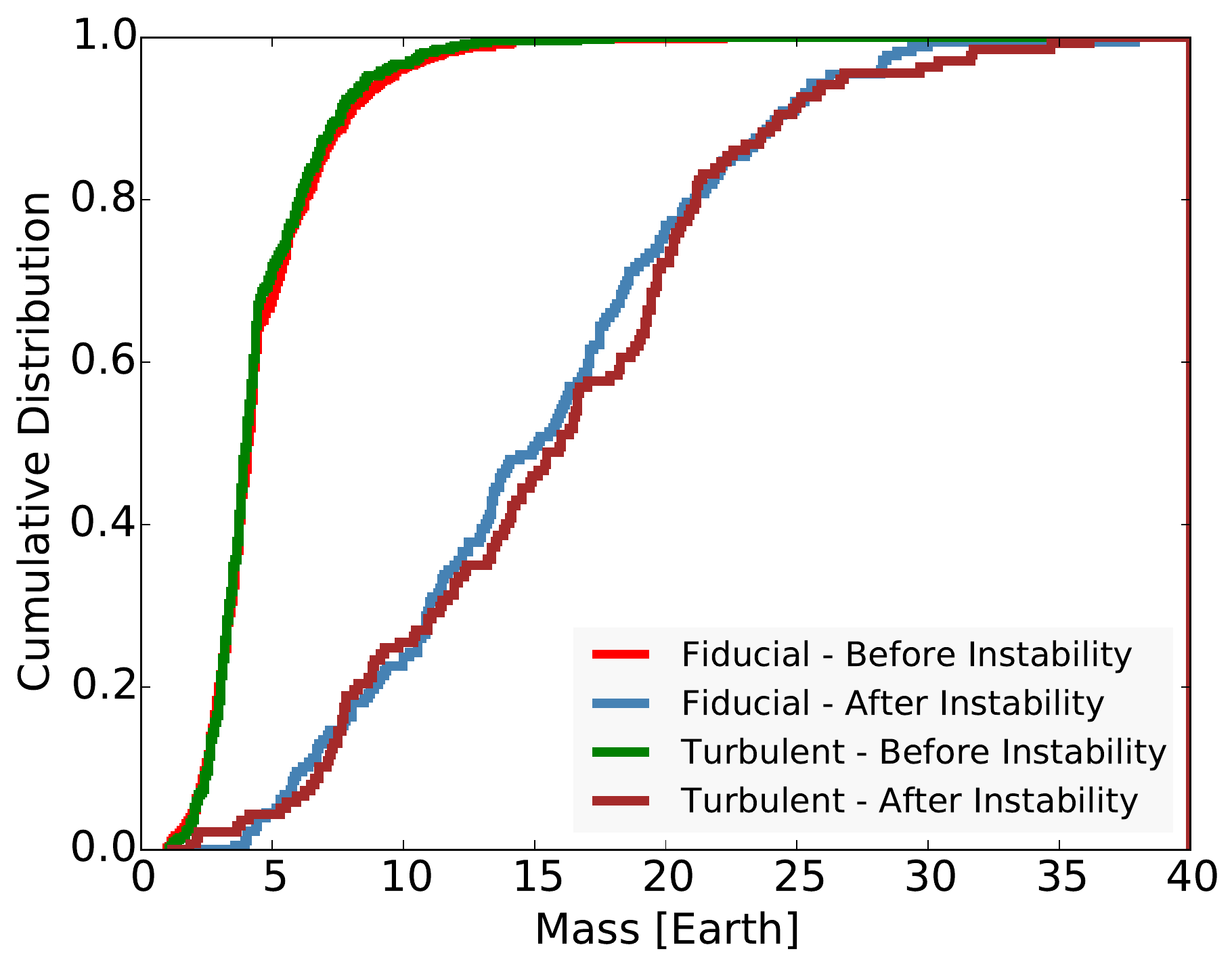}
     \includegraphics[width=.45\linewidth]{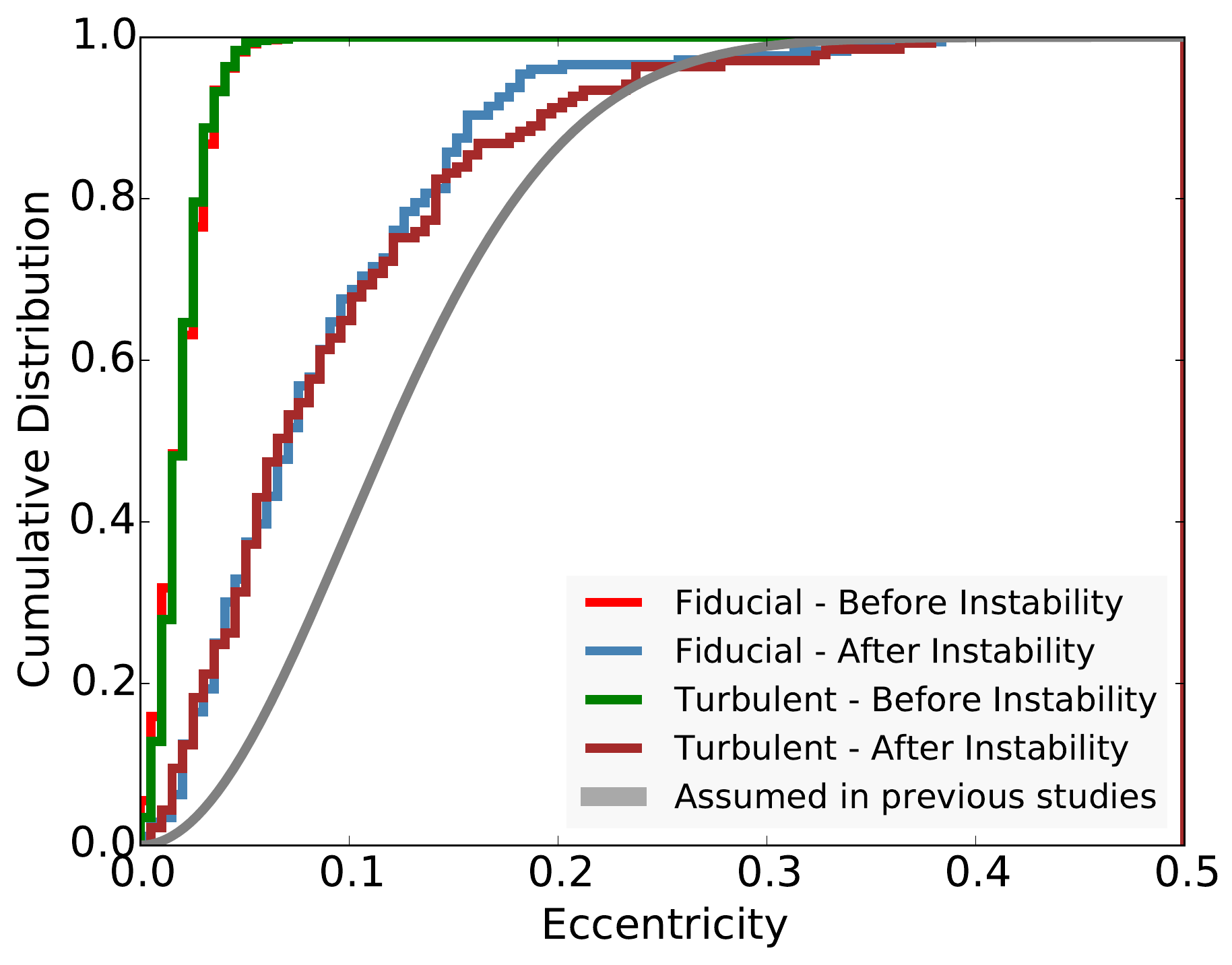}
      \includegraphics[width=.45\linewidth]{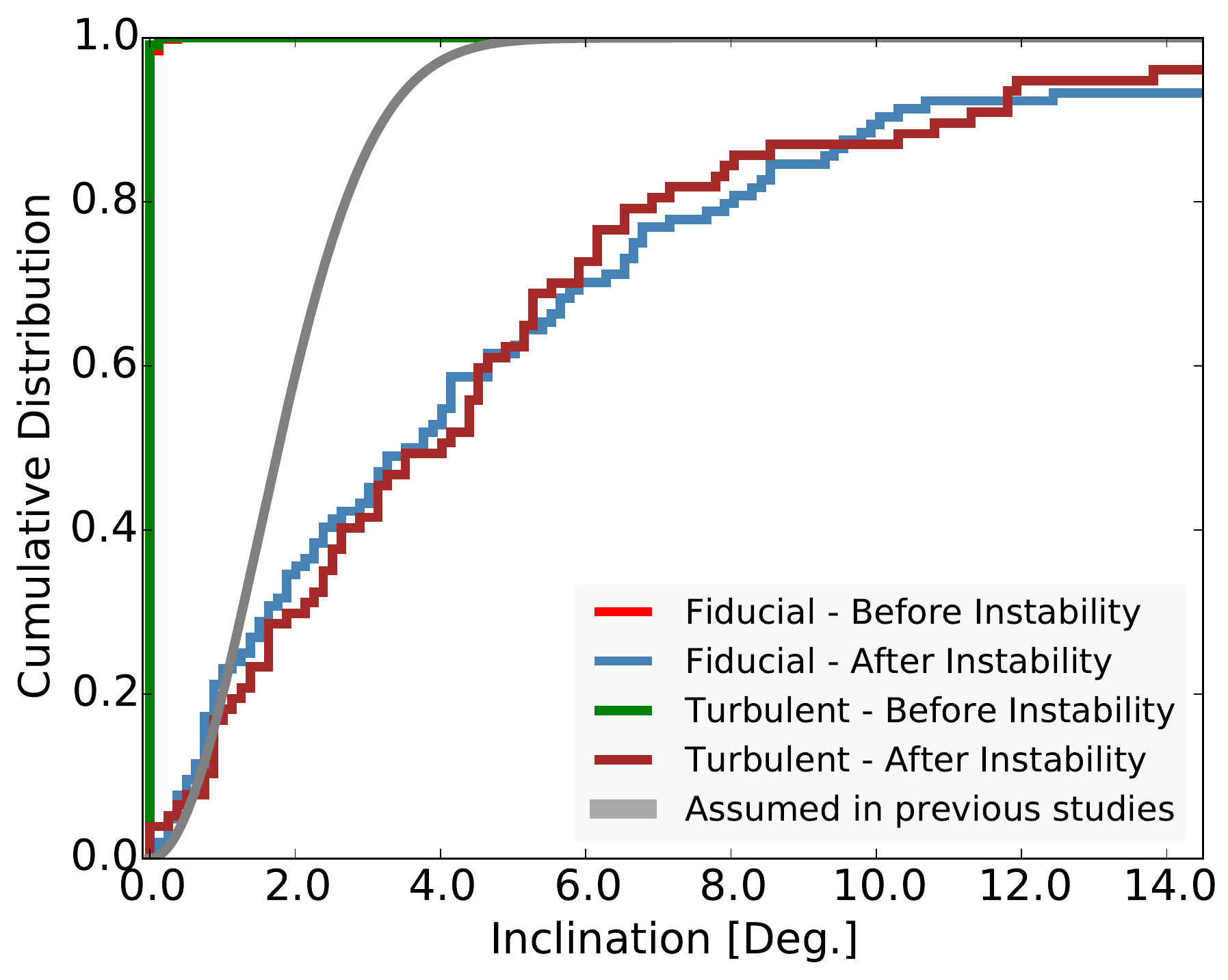}
    \caption{Cumulative distribution of semi-major axis ({\bf top-left}), mass ({\bf top-right}), orbital eccentricity ({\bf bottom-left}) and  mutual inclination ({\bf bottom-right}) of planets inside 0.5 AU in our simulations. The  eccentricity and orbital inclination distributions of  observed Kepler planets derived from statistical analysis are shown for comparison in  the respective lower panels. The eccentricity distribution shown by the gray line follows  a Rayleigh distribution with $\sigma_e=.1$ \citep{moorheadetal11} while the mutual inclination distribution (gray line) follows a Rayleigh distribution with $\sigma_i=1.5^\circ$ \citep{fangmargot12}. The latter put more than 85\% of the planets with orbital inclination smaller than 3 degrees.} 
    \label{fig:a_m_e_i}
\end{figure*}

Figure~\ref{fig:a_m_e_i} shows the cumulative distribution of semi-major axis, mass, orbital eccentricity and mutual inclination of simulated planets inside 0.5~AU. Cumulative distributions of semi-major axis (left-hand upper panel of Figure~\ref{fig:a_m_e_i}) are broadly identical before and after instabilities. Nevertheless, it should be natural to expect that, before dynamical instabilities, simulations have a much smaller fraction of planets far out than afterwards, where planets have been scattered everywhere. However, this effect is not evident in the cumulative distributions because it is normalized to account for planets only inside 0.5~AU. The only difference is that the, before instability, systems show a pileup of planets at about 0.15~AU, near the disk inner edge, whereas this edge has been wiped out in the after instability systems.  There is no significant difference between fiducial and turbulent simulations. 

The top right panel of Fig.~\ref{fig:a_m_e_i} shows the cumulative mass distributions. Planets before instability are far less massive than afterwards.  The median planet  mass after instability is about 3 times larger.  This is simply due to the fact that the unstable systems  underwent a late phase of collisions\footnote{ Note that, in the inner regions, collisions are much more common than ejections. It is an issue of Safronov number~\citep{safronov1972}.}. There is again no difference between turbulent and fiducial simulations. 

The bottom panels of Fig.~\ref{fig:a_m_e_i} show the eccentricity and mutual inclination distributions of simulated planets.  The orbital distribution of Kepler planets was derived in a series of statistical studies \citep{lissaueretal11a,kaneetal12,tremainedong12,figueiraetal12,fabryckyetal14,plavchanetal14,ballardjohnson14,vaneylenalbrechts15}. To represent the eccentricity distribution of observations derived from statistical studies we used a Rayleigh distribution with with $\sigma_e=.1$ (e.g. \cite{moorheadetal11}). For the inclination distribution we used a Rayleigh distribution with $\sigma_i=1.5^\circ$ (e.g. \cite{fangmargot12}).

The eccentricities and inclinations of planets before instability  (or equivalently of planets in stable systems) are extremely low.  These resonant chains have very low eccentricities and are close to perfectly coplanar.  As expected, the eccentricity and inclination distributions of the planets are strongly affected by dynamical instabilities.  Compared to the eccentricity distribution inferred from statistical analysis, planets  after instabilities are in better agreement with the assumed values but the difference is still noticeable (Fig.~\ref{fig:a_m_e_i}; bottom-left). Of course, we have used a single distribution to represent the expected quantities while the real data may require more than a single distribution to fit the data~\citep{lissaueretal11a}. The inclination distribution of planets in our simulations is also quite different from the suggested by previous studies (Fig.~\ref{fig:a_m_e_i}; bottom-right). In Section 6.2, we will discuss how the inclination distribution of planets in our simulations compare to the distribution inferred from  statistical analysis.

\section{Matching the observed Kepler planets}

In our simulations super-Earth systems follow a typical evolutionary path.  Planets grow, migrate inward and pile up into resonant chains far more compact than the observed ones (see Fig.~\ref{fig:periodratio_unstable}).  A substantial fraction of these resonant chains become unstable, causing their planetary systems to spread out dynamically.  

Following the previous section, we divide our simulations into two batches: those that did not undergo instabilities after the dissipation of the disk and those that did. Stable simulations remain in compact resonant chains whereas the unstable systems have undergone a late phase of accretion and spread out considerably. Note that for this analysis we are using both the fiducial set of simulations and the set that included turbulence, given that we could find no significant difference in outcomes of these simulations (Section 5).


\subsection{The period ratio distribution}
Figure~\ref{fig:periodratio_unstable} shows the period ratio distribution of adjacent planet pairs in unstable systems before and after dynamical instabilities. Before instabilities planetary systems are clearly far more compact than the observed Kepler systems.  Similarly to planet pairs of stable systems, before instability planet pairs  are locked in resonant chains, seen as vertical lines in Figure~\ref{fig:periodratio_unstable} (see Figure~\ref{fig:unstable_vs_stable} for stable systems).  After instabilities, the unstable systems have spread out compared with those that remain stable, and are no longer preferentially found in resonance. In fact, they are  are modestly more spread out than observed Kepler systems. 


We now test the effect of planet mass on the planets' orbital spacing.  We divide our simulations (unstable systems) into three groups by total mass $M_{tot} = M_1+M_2$,  where $M_1$ and $M_2$ are the masses of adjacent planets pairs. The low-mass planet pairs have $M_{tot}< 25 \mearth$, the high-mass pairs have $M_{tot}>34.5 \mearth$, and the medium-mass pairs lie in between.  These boundaries were chosen to put a roughly equal number of planet pairs in each bin.

\begin{figure*}
     \includegraphics[width=.45\linewidth]{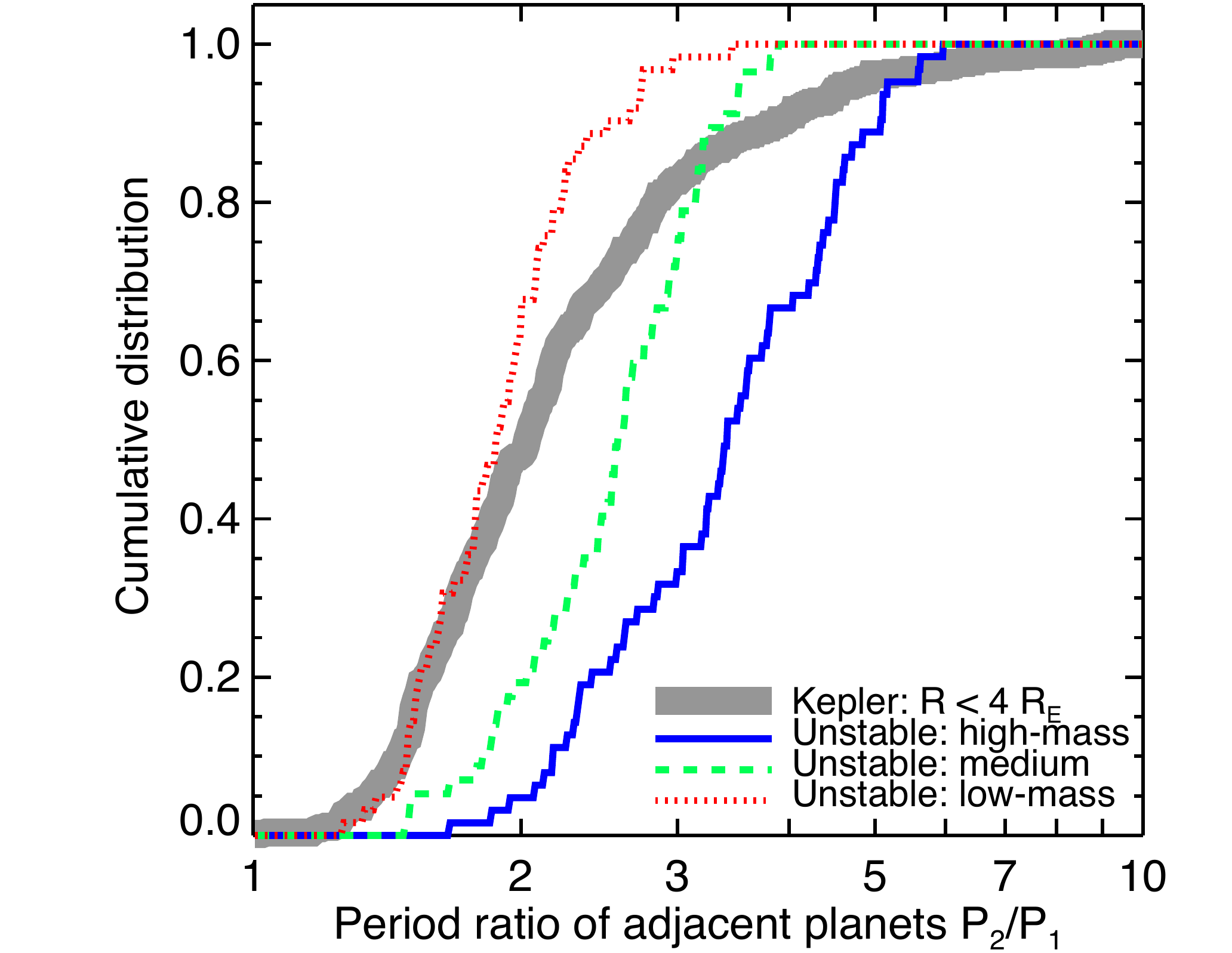}
     \includegraphics[width=.45\linewidth]{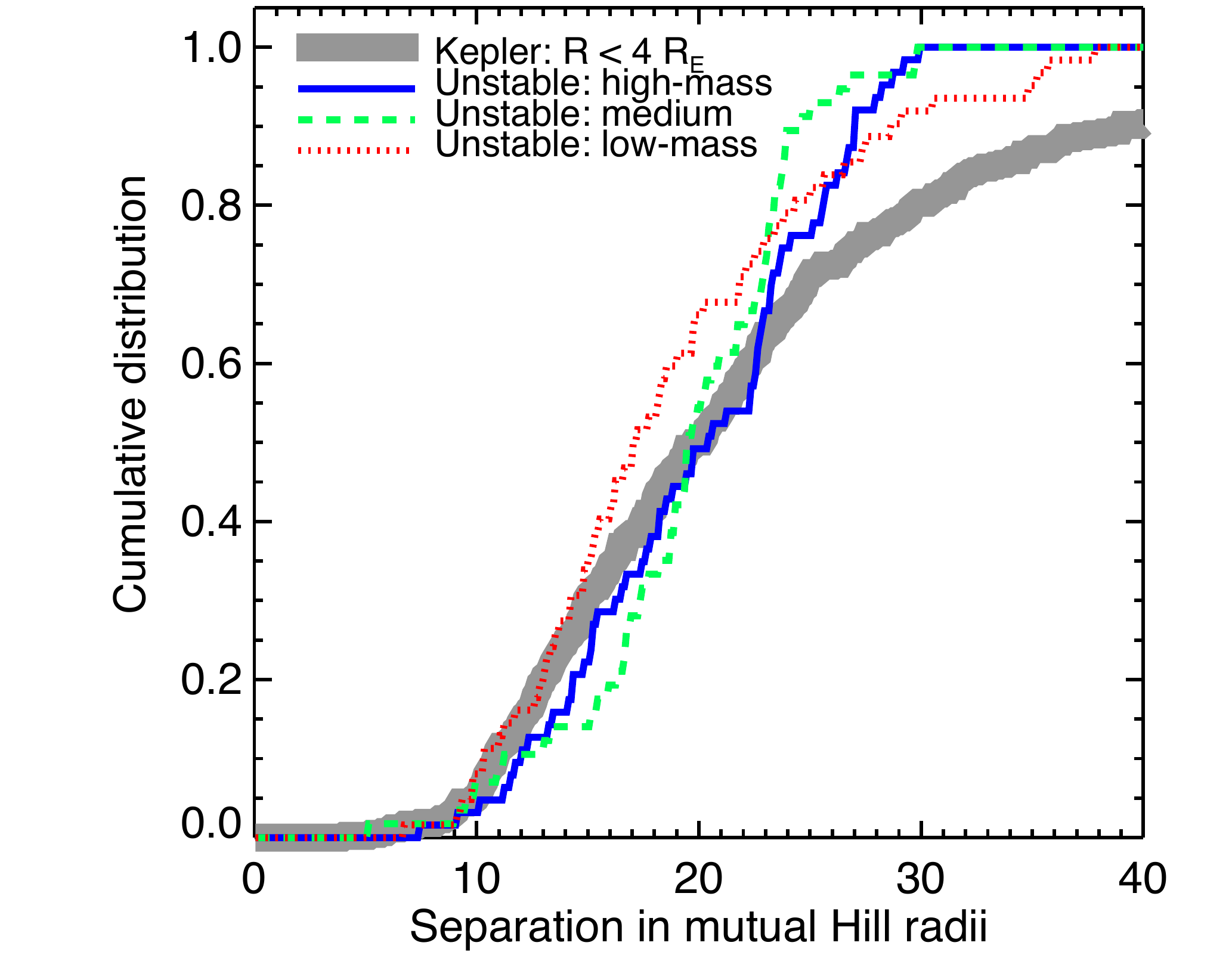}
    \caption{The separation of planet pairs in unstable simulations as a function of the total mass.  The low-mass planet pairs have $M_{tot} = M_1+M_2 < 25 \mearth$, the high-mass have $M_{tot}>34.5 \mearth$, and medium-mass pairs are in between.  {\bf Left:} Period ratio distribution.  {\bf Right:} Separation in units of mutual Hill radii.  Masses for the Kepler systems were attributed using the probabilistic mass-radius relation of~\protect\cite{wolfgangetal16}.}
    \label{fig:sep_mtot}
\end{figure*}

Figure~\ref{fig:sep_mtot} (right panel) shows that, after instabilities, planets are spaced by mutual Hill radii and not by period ratio.  Higher-mass planet pairs are systematically more widely-spaced than lower-mass pairs in terms of the period ratio of adjacent planets (left panel).  However, pairs of planets with different masses have very similar distributions when measured in mutual Hill radii.  They also provide a good match to the Kepler systems, for which we assigned masses using the probabilistic mass-radius relation of \cite{wolfgangetal16}:  $M = 2.7 \left(R/R_\oplus\right)^{1.3}$.  This fits nicely with the results of \cite{fangmargot12}, who inferred that  Kepler systems are typically spaced by roughly $20 \pm 10$ mutual Hill radii.  It also emphasizes that the Kepler systems appear to be the result of dynamical instabilities~\citep[see also][]{cossouetal14,puwu15}.

We have shown that the unstable planet pairs are more spread out than the observed ones, and that the stable planet pairs are more compact than the observed ones (see Fig.~\ref{fig:periodratio_unstable}). It follows that a mixture of the two populations may match observations.  Indeed, several chains of three or more resonant planets have been identified, such as Kepler-223~\citep{millsetal16}, Kepler-80~\citep{macdonaldetal16}, GJ 876~\citep{riveraetal10} and TRAPPIST-1 \citep{gillonetal17}.  While dynamical instabilities can generate resonances~\citep{raymondetal08b}, the delicate architecture of resonant chains indicate that they are signposts of stable systems.  Matching observations thus requires that a fraction of resonant chains remain stable after dissipation of the gas disk.  

We attempt to match the observed period ratio distribution using a mixture of stable and unstable systems from our simulations.  There are two challenges in this exercise.  First, our simulations provide planet masses but most Kepler observations only provide planetary radii.  Several studies have used mass constraints to produce mass-radius relationships for transiting planets (e.g. \cite{lissaueretal11b,fangmargot12,weissmarcy14}).  As above, we adopt the probabilistic study of \cite{wolfgangetal16}. We find that most of our simulated planets are significantly more massive than those inferred for the Kepler systems. As we impose a cutoff of $4 R_\oplus$ in the Kepler data, the maximum mass inferred using the Wolfgang mass-radius relationship is $16.4 \mearth$ and the median planet mass in the sample is $6.3 \mearth$.  The median mass of unstable planets at the end of our simulations is $14 \mearth$.\footnote{It's interesting to note that the median mass of the planets trapped in resonant chains is $5.6 \mearth$. This is comparable to the inferred value for Kepler sample of $6.3 \mearth$.  One might also speculate that the accretion of planets during late instabilities is inefficient due to collisional erosion and mass loss~\citep[see e.g.][]{leinhardtstewart12,Stewartetal12}.  That remains an interesting avenue for future study.}

The second challenge in this exercise is understanding observational bias. The transit probability scales linearly with the orbital radius such that close-in planets are far easier to detect~\citep[see][]{charbonneauetal07,winn10,wrightgaudi13}. For a perfectly coplanar planetary system aligned with the observer, every planet transits,  although it remains a higher probability than the outer planets will be missed.  But for a system with many planets on strongly inclined orbits relative to the observer, none transits. In a three planet system, if the middle planet does not transit but the inner and outer planets do, the inferred period ratio is $P_3/P_1$ rather than $P_2/P_1$ or $P_3/P_2$, and is pushed to a much higher value. This spreading to higher period ratios is a function of the mutual inclinations among planets.

We attempt to quantify observational selection effects by performing simulated observations of our planet pairs.  We wrote a simple code to observe each of our simulated systems from a large number of vantage points evenly spaced on the celestial sphere.  For a given line of sight we kept track of which planets were detected.  We then assembled the detected planetary systems from all line of sights.

\begin{figure*}
     \includegraphics[width=.45\linewidth]{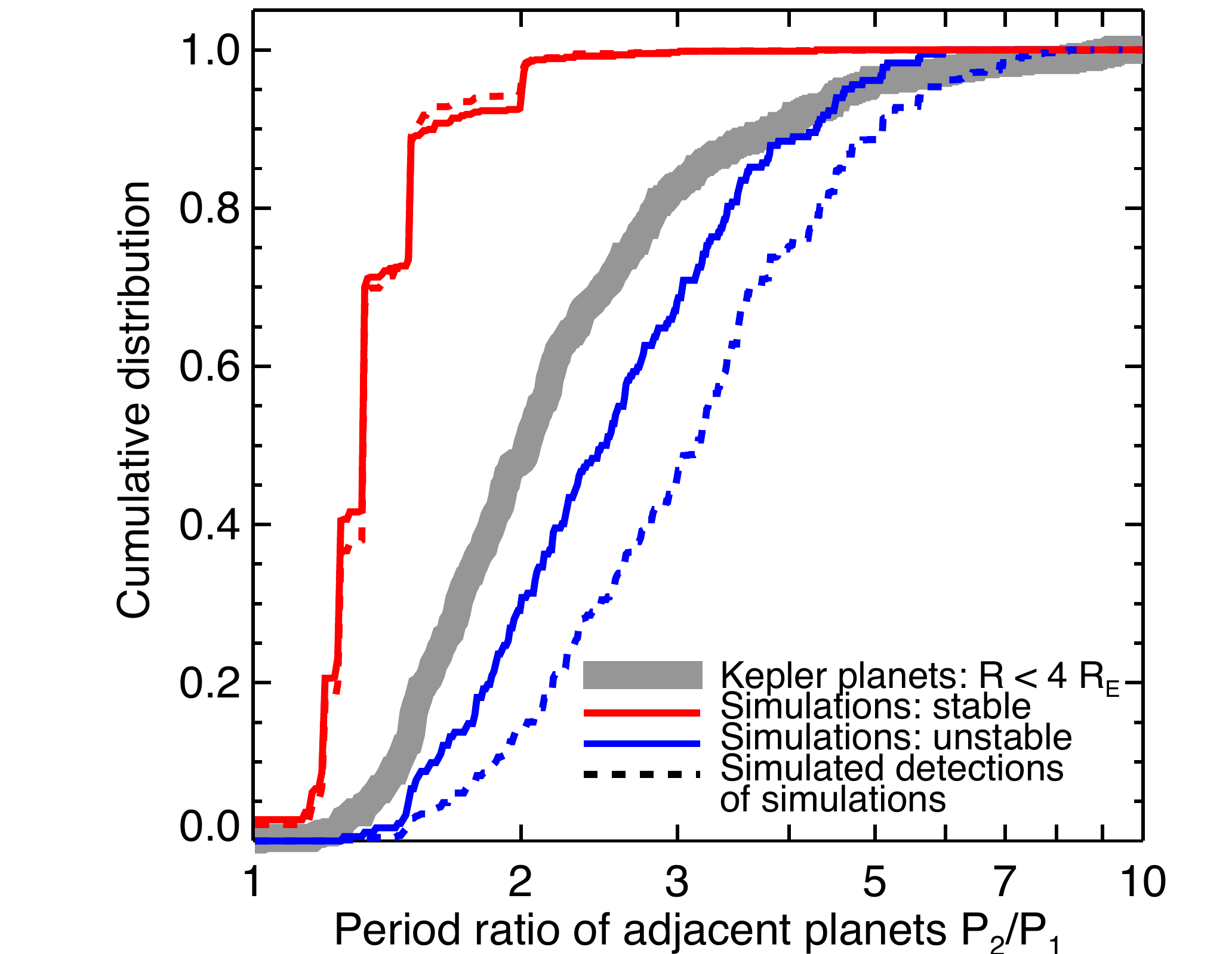}
     \includegraphics[width=.45\linewidth]{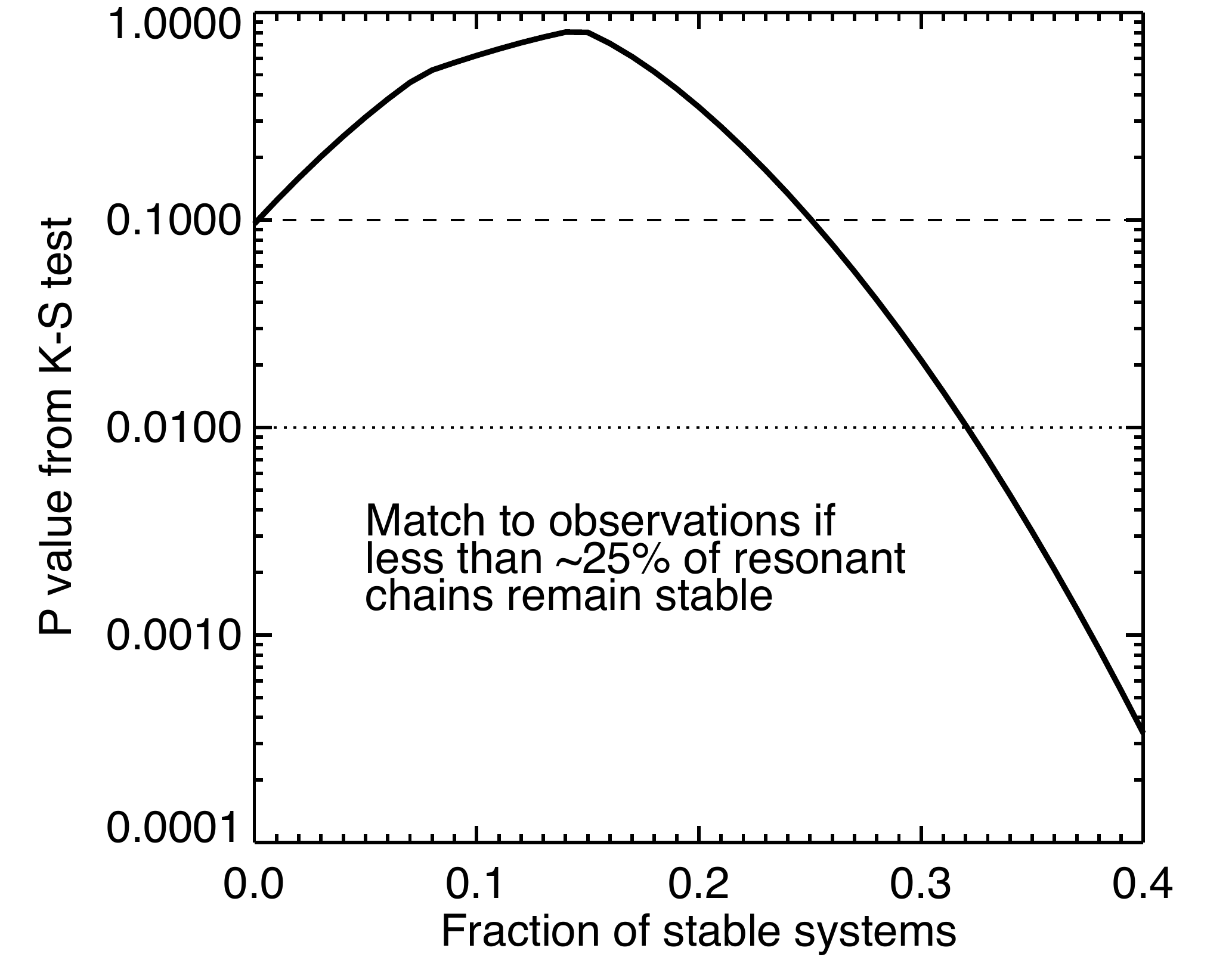}
    \caption{The outcome of our simulated observations.  {\bf Left panel:} The period ratio distribution of the observed Kepler systems (thick gray line), as well as our stable (solid red line) and unstable (solid blue line) simulations.  The dashed lines show the planet pairs retrieved by simulated observations.  {\bf Right panel:}        P values from K-S tests comparing the Kepler systems with a sample of simulated planet pairs with $M<16.4 \mearth$, after taking observational bias into account with simulated observations.  The dashed/dotted line is at $p=$10\%/1\%.} 
    \label{fig:Pratio-simobserv}
\end{figure*}

Figure~\ref{fig:Pratio-simobserv} (left panel) shows the effect of observational bias on our simulated systems.  The very low-inclination resonant stable systems were barely affected by observational bias.  The plane of these systems is so thin that they are almost always either detected or not (depending on the line-of-sight).  The higher-inclination unstable systems (e.g. compare Figures~\ref{fig:4panels_1} and~\ref{fig:4panels_2}) are more strongly affected by observational bias. Naturally, the orbital configuration of adjacent planet pairs in these systems are systematically shifted to larger period ratios, as planets are eventually missed in transit.

We performed a simple experiment to determine the best-fit combination of simulations to match observations.  We tested the effect of the mixing ratio of stable and unstable systems on how well the simulations match observations.  We restricted ourselves to planet pairs in which each planet was less massive than the Kepler cutoff of $16.4 \mearth$, which corresponds to the mass of a $4 R_\oplus$ planet with the \cite{wolfgangetal16} mass-radius relation. This is our upper size cutoff for Kepler planets. We also restricted this analysis to systems with $P_2/P_1 \le 3$, to reduce the contribution from systems with hard-to-characterize missed planets.  We then generated different samples of simulated planets by varying the fraction of stable systems $F$ included in the sample. We performed K-S tests to roughly judge the goodness of fit for each sample.  We note that even though our simulated observations contained thousands of planet pairs, the effective number of points used to generate $p$ values was limited by a combination of the number of simulated planet pairs with appropriate masses and the Kepler sample.

Figure~\ref{fig:Pratio-simobserv} (right panel) shows that our simulated observations roughly match the Kepler sample if less than 25\% of planet pairs come from stable simulations (for a probability $p \ge 10\%$ that the two samples are consistent with having been drawn from the same distribution).  This can be interpreted as an indicator of the fraction of observed systems that did not undergo an instability, for which the planets survived in a resonant chain.  The abundance of resonant chains among known Kepler systems is perhaps 5\% \citep{fabryckyetal14}.  We expect this to correspond to the contribution from stable systems.  Our simulations are indeed consistent with that value.

We expect that this analysis was further affected by the fact that our simulated planets were in general far more massive than the Kepler planets.  After an instability, the mutual inclinations between the planets' orbits naturally correlates with the strength of the gravitational scattering, i.e., the planets' masses.  Pairs of low-mass planets in systems with other, high-mass planets, may therefore have higher mutual inclinations than they otherwise would.  This ``inclination inflation'' should have the effect of pushing period ratios to higher values. We expect that modestly lower-mass systems would therefore have lower mutual inclinations compared with our simulations.  Lower-mass systems would appear more compact and thus require a smaller contribution of stable systems to match observations. 

To test the other extreme, we performed the same exercise as above without taking observational bias into account.  For the same restrictions as above ($M<16.4 \mearth$, $P_2/P_1 \le 3$), the best match to observations was for the smallest contribution from stable systems.  However, up to nearly 20\% of stable systems were allowed while maintaining $p \ge 10\%$.  

We conclude that our simulations can indeed provide an adequate match to the period ratio distribution of Kepler planet pairs.  We can firmly constrain the abundance of stable resonant chains to contribute less than $\sim 25\%$ of planet pairs.

\subsection{The Kepler dichotomy}

 The Kepler super-Earth sample is bimodal~\citep{lissaueretal11a,fabryckyetal14}: stars tend to either have one or many super-Earths~\citep{fangmargot12,ballardjohnson14}. It has been proposed that Kepler {\it dichotomy} is a signature of planet-planet scattering~\citep{johansenetal12}, in-situ growth close-in~\citep{moriartyballard16}, or instabilities produced by spin-orbit (mis-)alignment~\citep{spaldingbatygin16}.

We performed synthetic observations to determine whether our simulated planetary systems are consistent with the Kepler dichotomy.  We considered viewing angles spanning from 30 degrees above the initial $i=0$ plane to 30 degrees below, with even sampling in azimuth.  For each viewing angle we determined the number of planets that transited for each of our stable and unstable simulations. 

\begin{figure}
\includegraphics[width=.98\linewidth]{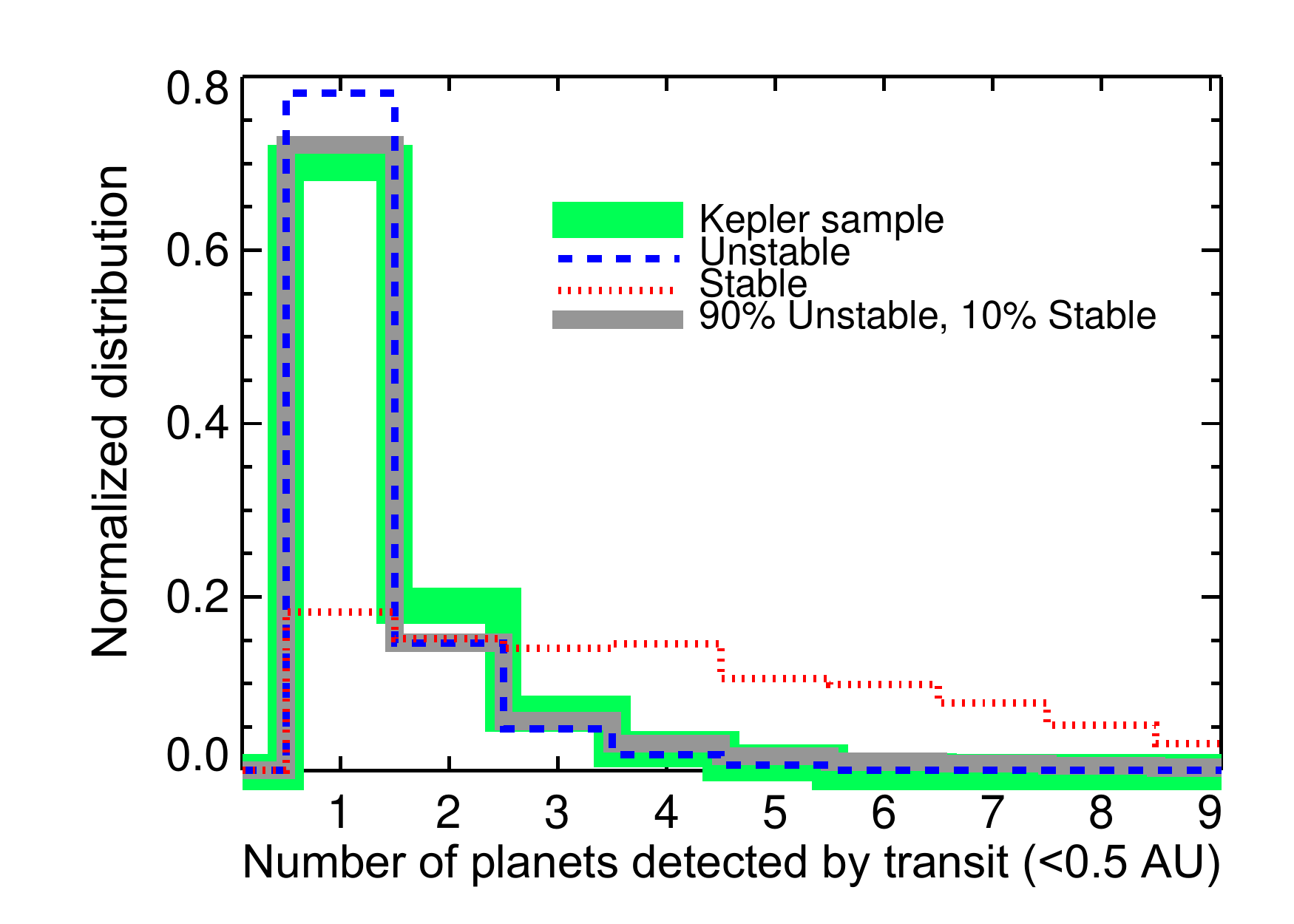}
\caption{The number of planets detected in synthetic observations of our simulated planetary systems.  The blue/red curves represent the unstable/stable simulations (combining the fiducial and turbulent sets).  The gray curve is a 90-10 mixture of the unstable and stable simulated systems, respectively.  The thick green curve is the Kepler sample, removing single giant planet systems but keeping systems with giants and super-Earths. }
\label{fig:dichotomy}
\end{figure}

From viewing angles where at least one transiting planet was detected, Figure~\ref{fig:dichotomy} shows the distribution of the number of transiting planets.  The difference between stable and unstable systems is striking.  Stable systems have such low mutual inclinations that it is common to detect high-N systems. Only 18\% of detections were of a single planet in transit, and 66\% of detections had $N \ge 3$.  In contrast, for unstable systems 78\% of outcomes were single-planet detections and only 7\% of detections had 3 or more planets. 

If we combine stable and unstable systems in a 1-to-9 ratio (i.e., with 90\% unstable and 10\% stable), we naturally obtain a ``dichotomy'' that is almost identical to the observed Kepler dichotomy (Fig.~\ref{fig:dichotomy}).  The significant mutual inclinations in the unstable systems produce a peak at $N=1$, while the very low mutual inclination stable systems provide a long tail to high-N.  Low-multiplicity systems $N=1-2$ are dominated by unstable systems whereas high-multiplicity systems ($N \ge 4$) are more often stable.  The Kepler-223~\citep{millsetal16} and TRAPPIST-1~\citep{gillonetal17} multi-resonant super-Earth systems appear to be good examples of high-N stable systems.

This would suggest that the Kepler dichotomy is simply an observational artifact.  Our simulated super-Earth systems naturally produce a spike of apparent singleton planets. However, each of those systems contains at least one -- and in some cases many more -- additional planets within 0.5 AU and beyond.  As we showed above this same sample of simulations matches the observed period ratio distribution (see Figure \ref{fig:Pratio-simobserv}). Of course, we do not claim to match all the details of the Kepler sample, but our results strongly suggest that there is no need to invoke special evolutionary histories for single super-Earth systems.  If, on the other hand, singleton super-Earth systems turn out to have a high false positive rate, this analysis would need to be re-visited.

 Previous studies had mixed success in matching the Kepler dichotomy.  \cite{johansenetal12} drew from the observed period ratio distribution and varyied the width of a Gaussian distribution of mutual inclinations.  Starting from triple-planet systems, they were unable to reproduce the dichotomy, quantified as the relative abundance of observed triple-, double-, and single-planet systems.  \cite{lissaueretal11a} and \cite{ballardjohnson14}  were equally unable to match the Kepler dichotomy simply because their single component-model of planetary architectures underestimates the number of single planet systems.

\begin{figure*}
\includegraphics[width=.49\linewidth]{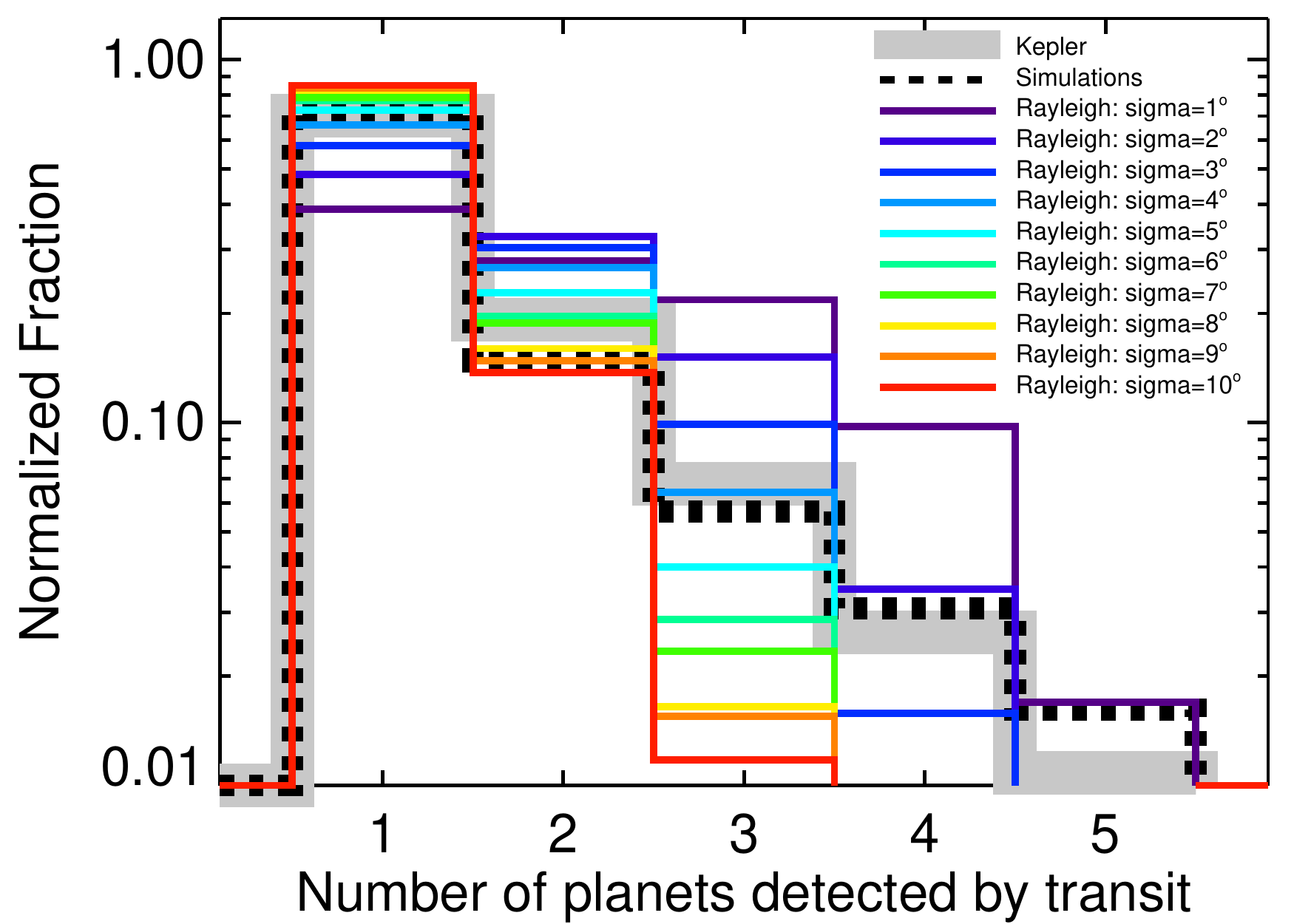}
\includegraphics[width=.49\linewidth]{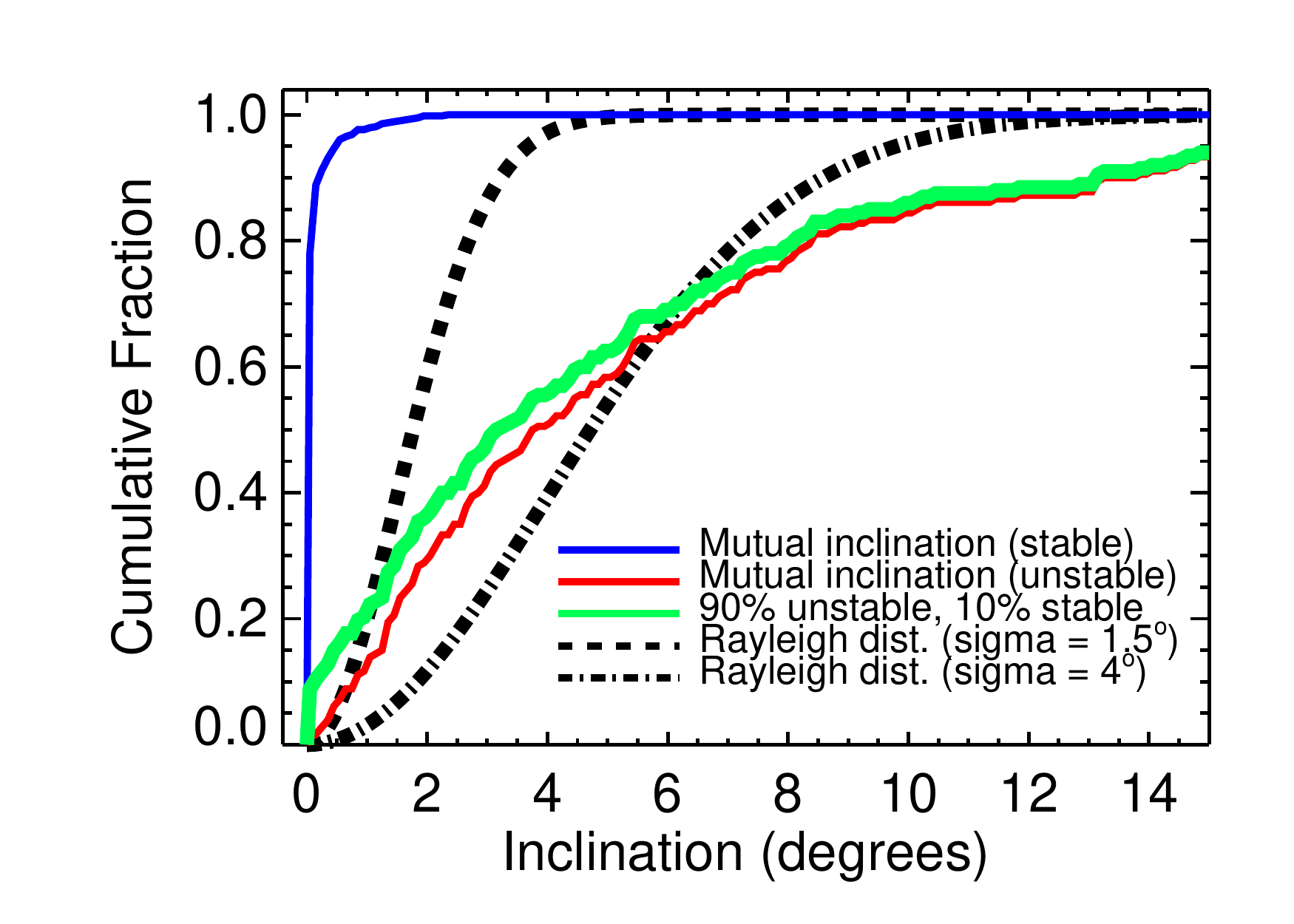}
\caption{ Comparison of the distribution of the number of planets in our simulations and statistical studies. {\bf Left:} Distribution of the number of planets of the observed Kepler systems (thick gray line), as well as our unstable  simulations (dashed line), and the distribution derived from different statistical distributions of mutual inclinations of planets.  {\bf Right:} Inclination distributions in our stable systems (blue line), as well as in all our unstable systems (red line), the mix of unstable of unstable systems (90\% unstables plus 10\% stables; green line), Rayleigh distribution with {\rm $\sigma=1.5^\circ$} \citep[dashed line; see][]{fangmargot12}, and  Rayleigh distribution with {\rm $\sigma = 4^\circ$} (dash-dotted line).}
\label{fig:multiplicity}
\end{figure*}

Building on the work of \cite{johansenetal12}, we tested whether we could match the dichotomy with a single inclination distribution (single component-model of system architecture).  We generated synthetic planetary systems as follows.  The closest planet was placed between 0.05 and 0.1 AU, and subsequent planets were spaced by drawing a period ratio evenly between 1.5 and 3. Systems extended out to 0.5 AU, naturally providing a wide range of planet multiplicities. The planets' orbital inclinations were drawn from a  Rayleigh distributions with $\sigma$ varying from 1 to 10 degrees. The results of this experiment matched qualitatively those of \cite{johansenetal12}.  Figure~\ref{fig:multiplicity} (left panel) shows that no Rayleigh distribution matches the Kepler multiplicity distribution. The best result is for {\rm $\sigma=4^\circ$}, which provides an acceptable match for systems with 1, 2, or 3 planets. However, the Rayleigh distribution dramatically underproduces systems with $N > 3$.

 The reason our simulations match the dichotomy can be inferred from their mutual inclination distribution (Figure~\ref{fig:multiplicity}, right panel). The small contribution of near-coplanar stable systems provides large $N$ systems whereas the broad inclination distribution of unstable systems creates a peak at low $N$.

 \cite{fangmargot12} were able to match the dichotomy by introducing an additional parameter: the multiplicity distribution. This extra parameter renders the problem much simpler.  For instance, consider perfectly co-planar planetary systems. The observed distribution can be retrieved in a straightforward way if the multiplicity distribution of these systems matches the observed distribution (to within a small observational bias). Indeed, assuming certain statistical distributions for the number of planets and mutual orbital inclinations of planet pairs,  \cite{fangmargot12} were able to match both the number of multiple and single-planet Kepler systems for some combinations of parameters.
 
 \cite{hansenmurray13} performed in-situ growth simulations starting from a set of initial conditions (planetary embryos) reflecting a putative radial distribution of mass in solids. Similar to previous studies, they found that the number of single planet systems are more common in the Kepler data than in their simulated population. In fact, to match the Kepler multiplicity distribution in-situ growth simulations also require a very specific mix of protoplanetary systems produced from simulations with quite different initial conditions. Simulations starting with a distribution of protoplanetary embryos derived from a shallow disk can account for the observed single planet population while simulations starting with sufficiently steep radial mass distributions of solids tend to produce multiple transiting planets. A mix of planetary systems produced from steep and shallow disks allows one to build a good match to the Kepler planet multiplicity \citep{moriartyballard16}. Indeed, the radial distribution of mass in solids in real protoplanetary disks may  vary from disk to disk but in-situ accretion simulations also predict that the radial distribution of planets in multi-planet systems should be mass ranked. More specifically, they predict that planet mass should decrease with semi-major axis \citep{ogiharaetal15a,moriartyballard16}. This expected signature seems to be at odds with the Kepler data.

 Our simulations match the observed Kepler dichotomy with no free parameters or ad-hoc assumptions. Even the relative abundance of stable and unstable systems was chosen to be the one that already matches the observed period ratio distribution. The mutual inclination distribution in our simulations is clearly not a simple function like the ones used in statistical studies.  We encourage statistical studies to connect with dynamical models to include more realistic inputs. 

 There is one significant difference between the population of planets that matches the dichotomy (Fig. \ref{fig:dichotomy}) and the one that matches the period ratio distribution (Fig. \ref{fig:Pratio-simobserv}).  In constructing our sample to match the period ratio distribution, we selected the lowest-mass planet pairs from our simulations because those were the most reasonable match to the observed planets.  However, in each simulated system there is generally a mix of planet pairs of different masses, so we could not select just the lower-mass pairs when performing simulated observations. However, the inclinations in unstable systems are generated by dynamical instabilities after gas dispersal. This is a mass-dependent process, as the degree of excitation scales with the planets' escape velocities.  We do find lower mutual inclinations for lower-mass planet pairs. For whole systems of lower-mass planets one might expect systematically lower mutual inclinations.  This would tend to create a smaller peak at $N=1$.  While we have not performed this exercise (because a new set of simulations would be required), we can infer the expected outcome.  We expect our simulations to be able to match the multiplicity distribution for $N \ge 2$ but underestimate the Kepler peak at $N=1$.  This would lead us to predict that the observed Kepler singletons include a certain percentage of planets that we cannot explain.  These might be false positives or perhaps true singletons generated by another mechanism~\citep[see, e.g., ][]{izidoroetal15a}.

\section{Comments on different disk turbulence effects in different works}

To compare how the effects of turbulence affect the outcome of our simulations we calculated a diversity of parameters at two different ages of our simulations. This was done first at the disappearance of the disk and then after 100 Myr of integration in a gas free scenario. We found that essentially there are no differences between the results of simulations assuming turbulent or non-turbulent disk, both at the end of the disk lifetime and in the aftermath of the planet instability.  Here we compare our results with other previously published in the literature.

 \cite{adamsetal08} studied the effects of disk turbulence in the context of gas giant exoplanets using a stochastic pendulum model. These authors conclude that resonances should be rare in turbulent systems. This result is different from what we find here since our fiducial and turbulent models produced very similar results. So let's interpret the origin of this difference. First of all, we have to recall that our study is dedicated to low mass planets while these authors have focused on the effects of disk turbulence for gas giant planets. Yet, we have used quite different approaches. For example, in stark contrast with our simulations, the model by \cite{adamsetal08} assumes that the planet pairs are initially already locked in mean motion resonance. More importantly, they neglect the effects of planet  migration, gas disk dispersal, tidal damping of orbital eccentricity and inclination while applying the turbulence forces and checking on whether the rogue resonant pair will survive in resonance or not.

\cite{rein12} numerically studied  the effect of stochastic migration of planets in systems extracted from Kepler data. As transit-surveys only provide the radius but not mass of these planets he used the mass-radius relationship of \cite{fabryckyetal12} to estimate the planetary masses. The stellar mass and planet periods were also taken from the Kepler Objects of Interest (KOI) catalog. Planets' eccentricities and orbital inclinations were set to zero. To mimic type-I migration and eccentricity damping \cite{rein12} used simple migration timescales of about $10^3$ to $10^4$ years. In his prescription for the stochastic forcing he assumed that the turbulent strength is a small fraction ($\sim 10^{-5}-10^{-6}$) of the gravitational force from the central  \cite{reinpapaloizou09}.

The main result in \cite{rein12} is that it is possible to reproduce the period ratio distribution of close-in super-Earths if the effect of stochastic forcing is included in the simulations during migration. However, the results of our simulations are very different from his. We can understand the differences from at least three sources. Let's first focus on the gas disk phase.

The first great difference between our model and Rein's is how synthetic forces were implemented.  In \cite{rein12}'s simulations only the outermost planet of the system is allowed to type-I migrate and feel the damping of eccentricity and inclination. Yet, in his model all planets feel stochastic forcing. The motivation for this choice is unclear. In our simulations all planets felt stochastic forcing and type-I, eccentricity and inclination damping self-consistently.  Another important issue is that Rein's simulations were only integrated for $10^4$ orbital periods of the outermost planet of the system. This integration time may not guarantee capture in resonance for a large fraction of the planet pairs in the Kepler catalog (see Figure 1 in \cite{rein12}). Basically, the approach in \cite{rein12} strongly favors the production of dynamically relaxed planetary systems.  While Rein's simulations had short integration time, our protoplanetary disks lived for about 5 Myr. Thus,  planets in our simulations had the time to migrate to the inner edge of the disk and pile up in long chain of resonances before the gas dissipated (e.g. Figure~\ref{fig:4panels_3}). In our simulations the disk surface density decreases with increasing time according to the disk model derived from hydrodynamical simulations \citep{bitschetal15}. Finally, the third reason which may explain the difference between the results of these models  comes from the stochastic forcing model. We used a more sophisticated prescription than \cite{rein12} to mimic the effects of turbulence in the disk of gas. For example, Rein's simulations scale the stochastic force using simply a fraction of the relative gravitational force from the central star while our model scales the strength of the turbulence taking into account the disk surface density, aspect ratio and the distance between the center of the gas fluctuation and the planet \citep{laughlinetal04,ogiharaetal07}.  The stochastic kicks that each planet feels in Rein's simulations are uncorrelated because their model is based on a first order Markov process. This means that even in the case when planets are very close to each other they may feel very different stochastic kicks \citep{reinpapaloizou09}.  Moreover, the amplitude of the turbulence forcing in his simulations is purposely chosen to produce results consistent with observations~\citep{batyginadams17}. The model of \cite{laughlinetal04} is more robust in this sense.

It may be easier to understand our results if we compare them with those in \cite{ogiharaetal07} where the stochastic forcing model is similar.  In our fiducial simulations we set $\alpha=5.4\times 10^{-3}$ (alpha-viscosity parameter), thus the turbulence strength parameter $\gamma$ is about $\sim 2.5\times 10^{-4}$ inside 1 AU (see Eq. 38). Recall that in our model $\gamma$ comes from \cite{baruteaulin10} where it is obtained from calibration with three-dimensional magnetohydrodynamics calculations. Still, the typical values of $\alpha$ in MRI active zones is in the range  of $10^{-3}$ and $10^{-1}$ \citep{fromangpapaloizou07} which is in great agreement with our chosen value. Curiously, the simulations in \cite{ogiharaetal07} show that the effects of turbulence were only pronounced  in cases where $\gamma\geq10^{-1}$. Smaller $\gamma$ produced results very similar to those where turbulence was not included. We confirm this result with our simulations. However, it is not clear if $\gamma\gtrsim 0.1$ would be consistent with results from MHD simulations of very turbulent disks.

 It is also interesting to note that our results seem to agree well with conclusions from a  very recent paper by \cite{batyginadams17} where is derived an analytic criterion for turbulent disruption of mean motion resonances. According to these authors --at the inner regions of the disk-- only planet pairs with mass ratios smaller than $(m_1 + m_2)/M_{\odot}\lesssim 10^{-5}$ should be susceptible to disruption of resonant configurations. Given that  about 90\% (50\%) of our planet pairs have combined masses larger than 6$\mearth$ (10$\mearth$), this fits nicely with our results.


\section{Why do so many resonant chains go unstable?}
 
Figure~\ref{fig:Pratio-simobserv} shows that our simulations can match the sample of Kepler planets if at least 75\% of resonant chains go unstable and at most 25\% remained stable.  Among the Kepler multiple-planet systems, only a handful have been characterized as resonant chains (e.g.~\cite{millsetal16}).  There is an additional excess of planet pairs just exterior to the 3:2 and 2:1 resonances~\citep{fabryckyetal14}.  All told, it appears that $\sim5\%$ of Kepler planet pairs are in resonance.  If we equate that occurrence rate with the probability of a given system remaining stable in our simulations, it follows that $\sim 5\%$ of resonant chains remain stable.  This yields an acceptable match to the Kepler period ratio distribution (Fig.~\ref{fig:Pratio-simobserv}).

Only 50-60\% of resonant chains were unstable in our fiducial and turbulent sets of simulations (Fig.~\ref{fig:last_coll}).  This is a far cry from the roughly 95\% required to match both the period ratio distribution and abundance of observed resonances in the Kepler systems.  

How can we explain the deficit of unstable systems in our simulations?  Our simulations are idealized and are missing several physical effects.  For instance, we have only used a crude description of the conditions at the inner edge of the disk, which is critical in anchoring resonant chains.  We did not include the effects of general relativity, which causes close-in orbits to precess.  We also did not include tidal interactions between the planets and the star.  As young stars may dissipate more strongly than previously thought~\citep{mathis15}, it is possible that tides play an important role (Bolmont et al, in preparation). For simplicity, in our simulations the inner edge of the disk remains fixed at $\sim$0.1AU during all disk lifetime. However, in reality, the inner edge of the disk should move as gas the disk and star evolve, due to the balance between the stellar magnetic pressure and the viscous torque of the disk \citep[e.g.][]{koenigl91}. \cite{liuetal17} showed that under certain conditions two planets can experience divergent migration as the disk inner edge moves outward. This mechanism could be also another important inductor of dynamical instabilities.

We have neither considered gas accretion onto planetary cores \citep[e.g.][]{ginzburgetal16}, nor mass loss during planetary collisions \citep[e.g.][]{inamdarschlichting16}, nor effects of fast-rotating stars \citep{spaldingbatygin16}. Also, we have not included a reservoir of small planetary bodies, as pebbles or planetesimals, in the disk. Left-over planetesimals in the disk could be another potential trigger of later dynamical instabilities in these systems   \citep{chatterjeeford15}. We also did not explore the effects of the gas disk viscosity, the initial distribution of planetary embryos and photo-evaporation in the disk. All these are certainly interesting routes for future studies.

We should also recall that because we overestimate the masses of the real Kepler super-Earths, our simulations may simply underestimate the rate of instabilities. It is also possible that our simulations were stopped prematurely.  The number of instabilities was still increasing at 100 Myr, even though the rate was decreasing.  It is not out of the question that within 5 Gyr all of our resonant chains would have become unstable.  Indeed, for a similar-aged parent star a system of close-in super-Earths is dynamically far older than the Earth orbiting the Sun, because the super-Earths have completed 1-2 orders of magnitude more orbital periods.  Of course, if resonant chains simply undergo later instabilities, then the abundance of resonances should decrease with the stellar age.  Kepler-223, the best-characterized resonant chain to date, is a relatively old star~\citep[age of $\sim6$ Gyr][]{millsetal16}.  However, more data is needed to test this idea.

Given the successes of our model, we consider this shortcoming to be important.  We believe that we are missing a trigger to explain why so many resonant chains go unstable.  



\section{Conclusions}

We have simulated the migration and growth of system of close-in super Earths in an evolving gaseous disk model~\citep{bitschetal15}. We found that  stochastic forcing from disk turbulence have no measurable effect on the growth of close-in super-Earths. 

We propose that systems of super-Earths follow a standard evolutionary path. Embryos grow in the outer disk and migrate inward due to torques from the gaseous disk (Figs 
\ref{fig:map}-\ref{fig:mapplanets2}).  When the first embryo reaches the inner edge of the disk its migration is stopped by the planet disk-edge interaction~\citep{massetetal06} and other embryos migrate into a resonant chain with up to 10 or more close-in planets (Figs.~\ref{fig:4panels_1}-\ref{fig:4panels_3}) .  The configuration of resonant chains are far more compact than the observed Kepler systems (Fig.~\ref{fig:periodratio_unstable}).  As the gas disk dissipates, about 50-60\% of our resonant chains become unstable and undergo a late phase of giant collisions.  This spreads their orbits out and spaces them by mutual Hill radii rather than by orbital period ratio (Fig.~\ref{fig:sep_mtot}, left panel).  Our simulations match the period ratio distribution of the observed Kepler planet pairs if 75-100\% of resonant chains go unstable.  Taking into account the abundance of resonances among Kepler planets~\citep{fabryckyetal14}, we expect that the true instability rate of resonant chains is roughly 95\%. Our results also suggest that the large number of detected single planet systems  is simply an observational effect due to the mutual orbital inclination of planets in Kepler systems. Thus, although planets have been  essentially observed either in single or multiple planet systems our results suggest that this does not necessarily imply the existence of any dichotomy in the architecture of planetary systems.

Finally, when comparing our simulations with the Kepler systems, we are left with a mystery: why does it appear that so many resonant chains go unstable?

\section*{Acknowledgements}

We thank the anonymous referee for his/her constructive comments that helped us to improve the manuscript. A. I, S.~N.~R., A. M., A. P. and F. H.  thank  the  Agence  Nationale  pour la Recherche for support via grant ANR-13-BS05-0003- 01  (project  MOJO). Computer time for this study was provided by the computing facilities MCIA (Mésocentre de Calcul Intensif Aquitain) of the Université de Bordeaux and of the Université de Pau et des Pays de l'Adour. A. I  thanks also partial financial support from FAPESP (Proc. 16/19556-7 and  16/12686-2) during the final preparation of this paper. B. B. thanks the Knut and Alice Wallenberg Foundation for their financial support.





\bibliographystyle{mnras}
\bibliography{library} 


\bsp	
\label{lastpage}
\end{document}